%% file: traces_arxiv.tex
\documentclass{article}
\usepackage[margin=1in]{geometry}
\usepackage{subcaption}
\usepackage{multirow}
\usepackage{multicol}
\usepackage{hhline}
\usepackage{float}
\usepackage{wrapfig}
\usepackage{listings}
\usepackage{amsthm, amsmath}
\usepackage{graphicx}
\usepackage{url}
\usepackage{hyperref}

\usepackage{color}
\definecolor{lightgray}{rgb}{.9,.9,.9}
\definecolor{darkgray}{rgb}{.4,.4,.4}
\definecolor{purple}{rgb}{0.65, 0.12, 0.82}

\lstdefinelanguage{Solidity}{
  keywords={contract, mapping, uint, return, returns, function},
  keywordstyle=\color{blue}\bfseries,
  identifierstyle=\color{black},
  sensitive=false,
  comment=[l]{//},
  morecomment=[s]{/*}{*/},
  commentstyle=\color{purple}\ttfamily,
  stringstyle=\color{red}\ttfamily,
  morestring=[b]',
  morestring=[b]"
}

\lstdefinelanguage{EVM}[x86masm]{Assembler}
{
  morekeywords={SSTORE, JUMPDEST, PUSH1, DUP1, DUP2, DUP5, MSTORE, SWAP1,
  KECCAK256, SSTORE, POP, ADD, SLOAD},
  keywordstyle=\color{blue}\ttfamily,
  identifierstyle=\color{black},
  sensitive=false,
}

\newcommand{\var}[1]{\text{\lstinline+#1+}}
\newcommand{\geth}{\texttt{geth~}}
\newcommand{\geths}{\texttt{geth}'s}
\newcommand{\sload}{\var{SLOAD}~}
\newcommand{\sstore}{\var{SSTORE}~}
\newcommand{\emul}{\var{MUL}~}
\newcommand{\ediv}{\var{DIV}~}

\begin{document}
\title{An Empirical Study of Speculative Concurrency in Ethereum Smart Contracts}
\author{Vikram Saraph and Maurice Herlihy}
\maketitle

\begin{abstract}

We use historical data to estimate the potential benefit of speculative
techniques for executing Ethereum smart contracts in parallel. 
We replay transaction traces of sampled blocks from the Ethereum blockchain over time,
using a simple speculative execution engine. In this engine, miners attempt to execute
all transactions in a block in parallel, rolling back those that cause data conflicts.
Aborted transactions are then executed sequentially. Validators execute
the same schedule as miners.

We find that our speculative technique yields estimated speed-ups starting at
about 8-fold in 2016, declining to about 2-fold at the end of 2017,
where speed-up is measured using either gas costs or instruction counts.
We also observe that a small set of contracts are responsible for many
data conflicts resulting from speculative concurrent execution.
\end{abstract}

\input{intro}

\input{related}

\input{contracts}

\input{baseline}

\input{alternative}

\input{discussion}

\input{conclusion}

\bibliographystyle{abbrv}
\bibliography{blockchain}

\end{document}

%% file: intro.tex
\section{Introduction}
A \emph{blockchain} is a distributed data structure that implements a \emph{ledger}:
a tamper-proof, widely-accessible, append-only sequence of \emph{transactions}.
Blockchains form the basis for cryptocurrencies~\cite{ethereum,cardano,stellar,bitcoin,ripple}
and other applications that must maintain a shared state in the
absence of a trusted central authority.
In the Ethereum~\cite{ethereum} blockchain,
for example, if Alice wants to send a coin to Bob,
she broadcasts her transaction to one or more \emph{miners},
who package transactions into \emph{blocks},
and then undertake a consensus protocol to agree on which block
should be appended next to the shared blockchain.
A \emph{validator} is any party who reads the blockchain state and
checks it for correctness.
Miners are validators, of course,
but so is any party who needs to query the blockchain state.

In many blockchain systems,
client transactions can invoke scripts,
often called \emph{smart contracts}, or just \emph{contracts},
that perform logic needed to support complex services
such as trading, voting, and managing tokens.
Here, we focus on Ethereum-style smart contracts.

Ethereum's smart contracts present a concurrency challenge.
To reconstruct the blockchain's current state,
each validator must re-execute,
in a sequential, one-at-a-time order,
every call to every smart contract.
Such sequential validation is unattractive because it fails to exploit
the concurrency provided by modern multicore architectures.
Simply executing those calls in parallel is unsafe,
because there may be dependencies between contracts:
if one contract depends on the results of another,
then those contracts must be executed in the same order by every validator.
Because the Ethereum smart contract language is Turing-complete,
and because contracts can reference one another through untyped function pointers,
static analysis is unlikely to be broadly effective.

The most promising approach to concurrent execution of smart contracts
is \emph{speculation}~\cite{AnjanaKPRS2018,HyperledgerFabric,DickersonGHK2017}:
the virtual machine executes contract calls in parallel against the current state,
tracking each transaction's \emph{read set} and \emph{write set}
(memory locations read and written).
Writes to memory are intercepted and buffered.
Two transactions \emph{conflict} if they access the same memory
location, and one access is a write.
For every pair of conflicting transactions,
one is discarded, and the other is committed.
Speculative techniques typically work well when conflicts are rare,
but perform poorly when conflicts are common.

\subsection{Contributions}

How well does speculative concurrency work for smart contract execution?
This paper makes the following contributions.
We exploit publicly-available historical data to estimate conflict
rates in an existing blockchain.
This methodology,
replaying the historical transaction record against proposed
alternative run-times,
could be a useful model for other blockchain-centered investigations.
This study is \emph{exploratory}:
it aspires to provide a relatively fast and cheap estimate of how well
certain strategies are likely to do in practice,
with the goal of focusing future research attention in directions more
likely to be productive.
Of course,
an exploratory study necessarily employs sampling, estimation,
and approximation.

As described below in more detail,
we re-execute blocks sampled from the Ethereum blockchain
against a simple speculative execution engine.
This engine has two phases:
in the first (concurrent) phase,
all transactions are run in parallel.
In the second (sequential) phase,
transactions observed to conflict are discarded and re-run in
one-at-a-time order.
This execution strategy produced speed-ups ranging from about 8-fold
for blocks sampled from 2016,
gradually declining to about 2-fold for blocks sampled from 2017.

This study makes the following observations.
\begin{itemize}
\item 
  Even simple speculative strategies yield non-trivial speed-ups.
\item
  Over time, however,
  these speed-ups declined as transaction traffic increased.
\item
  Distinguishing between reads and writes is important:
  treating a transaction's read and write sets as a single conflict set
  substantially increases conflict rates.
\item
  More aggressive speculative strategies,
  such as running multiple concurrent phases,
  yield little additional benefit.
\item
  Accurate static conflict analysis may yield a modest benefit.
\item
  Increasing the number of cores in the simulated virtual machine from 16 to 64
  improved speed-ups, but there was little improvement above 64 cores.
\item
  In high-contention periods,
  most contention resulted from a very small number of popular contracts.
\end{itemize}
These observations suggest some directions for further research.
\begin{itemize}
\item In periods of high contention,
  most conflict is caused by a small number of very popular contracts.
  Today, contract writers have no motivation for avoiding such conflicts.
  It could be productive to devise incentives,
  perhaps in the form of reduced gas prices,
  for contracts that produce fewer data conflicts.

\item
  Many data conflicts, such as crediting and debiting account balances,
  are probably artifacts of defining conflict na\"{i}vely in terms of read-write sets.
  Perhaps conflicts could be reduced by extending the virtual machine
  to provide explicit support for common commutative operations such
  as credits and debits.
\end{itemize}

\subsection{Methodology in Brief}
We replay each transaction in a block,
computing each transaction's read and write sets.
We then greedily sort the transactions into two bins:
the \emph{concurrent bin} holds transactions that do not conflict with
any other transaction already in the concurrent bin,
and the \emph{sequential bin} holds the rest.

We then estimate the elapsed time required to
(1) execute the concurrent bin transactions in parallel
(including the cost of detecting and discarding conflicting transactions),
followed by (2) sequentially executing the sequential bin transactions.

Since we do not have a parallel EVM implementation to test,
we estimate a transaction's running time in two ways:
either by the \emph{gas} it consumed,
or by the number of Ethereum Virtual Machine (EVM) bytecode instructions executed.
(Both measures are easy to compute, and yield similar results.)
The speed-up is the ratio between the estimated elapsed times
for the sequential executions versus
the longest speculative execution.

In the next section we describe related work. In Section 3, we outline Ethereum's architecture, smart contracts, and all relevant terminology. Section 4 describes the setup of our empirical study along with statistics summarizing observed results, while in Section 5, we consider various alternatives to the baseline setup. Finally, in Sections 6 and 7, we discuss conclusions and potential future directions for extending this work.

%% file: related.tex
\section{Related Work}
Smart contracts were first proposed by Szabo~\cite{Szabo1997}.

Bitcoin~\cite{bitcoin} includes a scripting language of limited power.
Ethereum~\cite{ethereum} is perhaps the most widely used smart
contract platform, running on a quasi-Turing-complete virtual machine.
Solidity~\cite{solidity} is the most popular programming language for
the Ethereum virtual machine.
Other blockchains that support smart contracts include
Corda~\cite{corda-contract} and Cardano~\cite{cardano-contract}.

Hyperledger Fabric~\cite{HyperledgerFabric} is a permissioned blockchain where
transactions (calls to smart contracts) are executed speculatively in
parallel against the latest committed state.
Transactions' read and write sets are recorded and compared,
and conflicting contracts are discarded.

Dickerson et al.~\cite{DickersonGHK2017} have proposed a speculative
execution model where miners dynamically construct a fork-join
schedule that allows concurrent executions without violating
transaction dependencies.
Anjana \emph{et al.}~\cite{AnjanaKPRS2018} propose a way to extend this
approach to lock-free executions.

%% file: contracts.tex
\section {Ethereum Smart Contracts}

In this section,
we sketch Ethereum smart contracts, mining, and validation of Ethereum blocks.

\subsection{The Architecture}

In Ethereum, as in other blockchains,
multiple \emph{nodes} follow a common protocol in which \emph{transactions}
from \emph{clients} are packaged into \emph{blocks},
and nodes use a consensus protocol to agree on successive blocks.
Each block includes a cryptographic hash of its predecessor,
making it difficult to tamper with the ledger.

Each client has ownership of one or more \emph{accounts}, so that each transaction
 occurs between a \emph{sender} account and a \emph{recipient} account.
The majority of transactions are one of two kinds: either a \emph{value transfer}, which 
is a purely monetary transfer of \emph{ether} from sender to recipient, or a
\emph{contract call}, where the sender account makes a call to code associated
with the recipient account.

Some Ethereum accounts, in addition to maintaining a balance of ether,
possess associated code called a \emph{smart contract}. A smart contract resembles an
object in a programming language, with a long-lived \emph{state} recorded in the blockchain.
This state is manipulated by a set of \emph{functions} called either directly by clients (top-level calls) or indirectly
by other smart contracts (internal calls). To ensure that function calls terminate, each computational step incurs a
cost in \emph{gas}, paid by the caller. The caller specifies a maximum amount of gas it is willing to pay,
and if the charge exceeds that sum, the computation is terminated and rolled back, and the caller's gas is not refunded.
Nevertheless, the rolled-back transaction is still recorded on the blockchain.

A smart contract's code consists of a sequence of bytecode instructions, taken from the Ethereum bytecode 
instruction set. Every bytecode instruction consumes a certain amount of gas.
Each client runs an instance of the Ethereum virtual machine (EVM), which
executes calls to smart contracts and runs their sequence of instructions. While users may author smart contracts in higher level
languages such as Solidity, these contracts must ultimately be compiled into EVM bytecode, since it
is the bytecode that is published in the blockchain. The virtual machine specification, the bytecode instruction set,
and all associated gas costs are described in Ethereum's ``Yellow Paper'' \cite{wood2014ethereum}.

\begin{figure}
\begin{subfigure}{.6\textwidth}
\begin{lstlisting}[language=Solidity]
contract StorageInterface {
    mapping (uint => uint) storage;
    
    function getValue(uint key) returns (uint) {
        return storage[key];
    }
    
    function setValue(uint key, uint value) {
        storage[key] = value;
    }
}
\end{lstlisting}
\caption{This code is a simple smart contract written in the Solidity language.
The contract, \var{StorageInterface}, contains a Solidity mapping named
\var{storage}, and has functions \var{getValue} and \var{setValue}. Mappings are essentially
hash tables that store key-value pairs, and are the primary means of accessing Ethereum contract
storage. Figure \ref{fig:bytecode} shows the a snippet of the bytecode after compiling the Solidity
contract. The bytecode is what is published in the blockchain.}
\end{subfigure} \hfill
\begin{subfigure}{.3\textwidth}
\begin{lstlisting}[language=EVM, firstnumber=247]
...
JUMPDEST
DUP1
PUSH1 0x0
DUP1
DUP5
DUP2
MSTORE 
PUSH1 0x20
ADD
SWAP1
DUP2
MSTORE
PUSH1 0x20
ADD
PUSH1 0x0
KECCAK256
DUP2
SWAP1
SSTORE
POP
...

\end{lstlisting}
\caption{Bytecode resulting from compiling the contract.}
\label{fig:bytecode}
\end{subfigure}
\caption{An example Solidity smart contract and a fragment of the corresponding bytecode.}
\end{figure}

\subsection{Mining and Validation}

Smart contracts are first executed by \emph{miners}, or
nodes that repeatedly propose new blocks to append to the blockchain.
When a miner creates a block,
it selects a sequence of client transactions
and executes their smart contract codes in sequence,
transforming the old contract state into a new state.

Once a block has been appended to the blockchain,
that block's smart contracts are re-executed by \emph{validators}, or
nodes that reconstruct (and check) the current blockchain state.
Each miner validates blocks proposed by other miners,
and older blocks are validated by newly-joined miners,
or by clients querying the contract state. Once a contract is in a block,
it is effectively re-executed forever (in Ethereum),
so contract executions by validators vastly exceed executions by miners.

As noted earlier, one drawback of the Ethereum protocol is that
a block's contracts are executed in a one-at-a-time order,
so miners and validators cannot exploit modern multicore architectures.
Contracts cannot be executed concurrently in a na\"{\i}ve way,
because they share storage, and may be subject to \emph{data conflicts},
that is, concurrent accesses to the same the storage variables.

In the absence of explicit concurrency guidelines,
we execute transactions \emph{speculatively} in parallel,
allowing non-conflicting contracts to commit,
but rolling back conflicting transactions,
and running them sequentially in a second phase.

A novel aspect of this study is that we analyze the effectiveness of
speculation against the \emph{historical} record of transactions actually executed
on the Ethereum blockchain, replaying their bytecode instructions.
We are not aware of an Ethereum virtual machine implementation that supports concurrency,
so we simulate concurrent transaction execution by stepping through each transaction's instructions,
using eager conflict detection to sort each block's transactions into
a conflict-free parallel bin, and a conflicted sequential bin.
This strategy is simple and scalable,
a natural starting point for an empirical study.

%% file: baseline.tex
\section{The Baseline Experiment}
Here we describe the baseline experiment testing the effectiveness of a
simple speculative execution strategy against historical transaction data from
the Ethereum blockchain.
In Section \ref{alternative},
we describe variations on this baseline strategy that probe the
sensitivity of our measurements,
as well as alternative speculative strategies.

\subsection{Setting up the Experiment}
We set up an Ethereum node by installing and running the Ethereum Go client,
also known as \geth\!\!.
The \geth client allows one to synchronize with other network nodes 
and reconstruct the blockchain by validating each block.
The client comes packaged with an interface for fetching data
from the blockchain,
as well as debugging tools for inspecting transactions.
The debugging API includes a utility for reproducing the bytecode
trace of a given transaction.

The \geth client can run in various synchronization modes,
which determine what and how much old state the client records.
For example, a client running in \emph{light} mode will keep track
of the blockchain's current state only,
while in \emph{archive} mode,
the client maintains the full log of all previous states. 

Using the \geth API,
it is straightforward to retrieve any given smart contract's bytecode.
However, the only way to tell how that bytecode affects the blockchain
state is by re-executing each transaction before it.
Therefore we use \geths~ tracing utility to re-execute transactions, running the
client in archive mode so it can reference past states.
We found that reconstructing the blockchain from scratch in this way
was much too slow to be feasible on an ordinary computer
(an archive blockchain sync could not even keep up with the current
blockchain growth rate),
so we obtained a baseline copy of the blockchain from the
ConSenSys archive~\cite{consensys} dating from Ethereum's origins in 2016
to early October 2017.
Starting from this base, we used the \geth utilities to synchronize up to December 2017.

\begin{figure}
\centering
\includegraphics[scale=0.5]{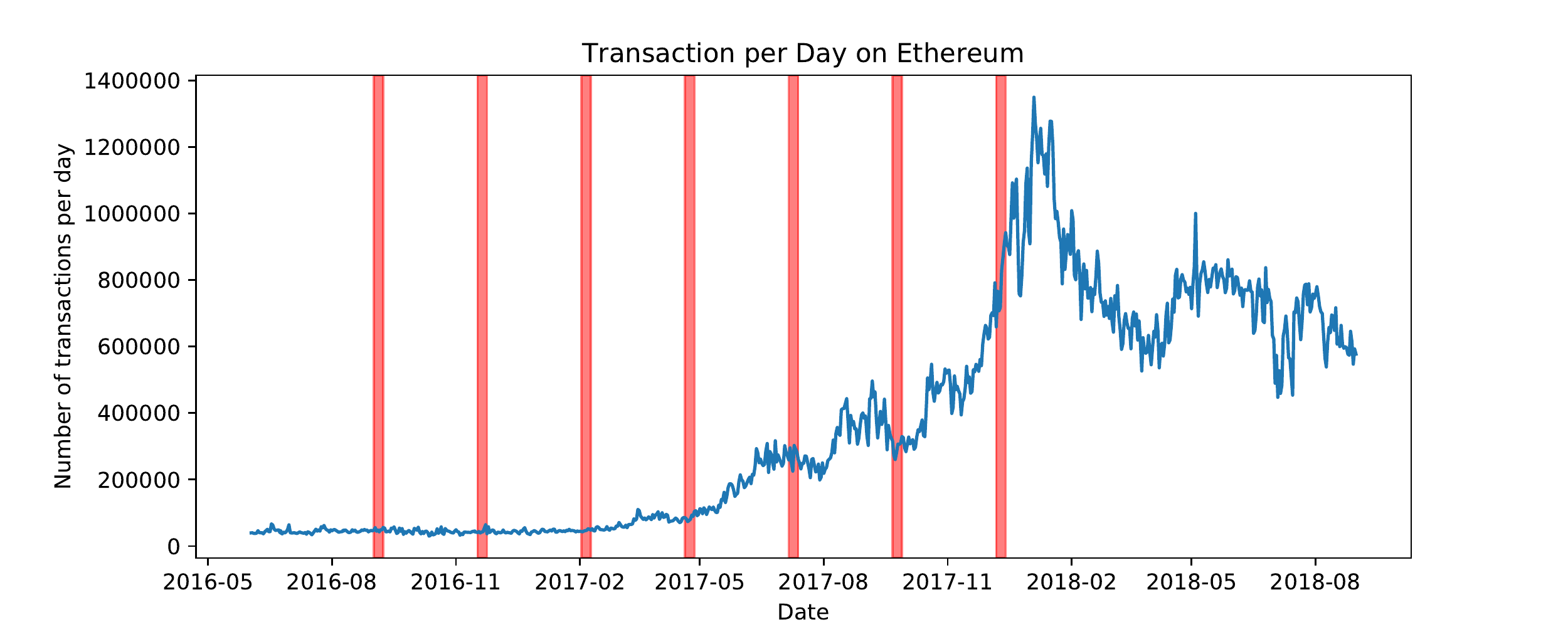}
\caption{Historical periods analyzed superimposed on number of transactions per day}
\label{intervals}
\end{figure}

Between these dates,
the Ethereum blockchain contains about 4 million blocks and 100 million transactions,
far too much data to analyze in detail in a reasonable time with reasonable
resources.
Instead,
we chose to focus on seven historical periods between July 2016
(after the DAO fork) and December 2017.
Each period spans roughly one week,
with consecutive periods separated by 11 weeks,
so that each day of the month is considered.   
Due to computational limitations,
we analyze every 10th block in each of the seven historical periods.
Figure \ref{intervals} shows these periods superimposed on the number of
transactions per day at that time. 

\begin{figure}
\centering
\begin{tabular}{|c|c|l|l|l|}
\hline
Period & Interval & Blocks & Transactions & Calls \\ \hline
1 & Sept 1 -- 8, 2016 & 4200 & 32502 & 11779 \\ \hline
2 & Nov 17 -- 24, 2016 & 4200 & 32648 & 11070 \\ \hline
3 & Feb 2 -- 9, 2017 & 3600 & 26415 & 10586 \\ \hline
4 & Apr 20 -- 27, 2017 & 4100 & 56051 & 23684 \\ \hline
5 & July 6 -- 13, 2017 & 3400 & 197003 & 84219 \\ \hline
6 & Sept 21 -- 28, 2017 & 2100 & 200980 & 96043 \\ \hline
7 & Dec 7 -- 14, 2017 & 4100 & 557471 & 255872 \\ \hline
\end{tabular}
\caption{Number of blocks, transactions, and contract calls analyzed in each historical period}
\end{figure}

\subsection{The Greedy Concurrent EVM}
We simulated a concurrent
EVM that executes transactions speculatively in parallel using the
following \emph{greedy} strategy.  For each block, execution proceeds
in two phases, an initial \emph{concurrent phase}, and a subsequent
\emph{sequential phase}.  We consider execution engines with either
16, 32, or 64 threads.  In the concurrent phase, each thread chooses a
transaction from the block and executes it speculatively.  If that
transaction encounters a conflict, the transaction's effects are
rolled back, and that transaction is deferred to the second,
sequential phase.  When a thread finishes executing a transaction, it
picks another to execute, continuing until all transactions in the
current block have been chosen.

The second phase starts when the first phase is complete:
the transactions that encountered data conflicts in the first phase
are re-executed sequentially. In the second phase, data conflicts
are not an issue since transactions are explicitly serialized and
state changes committed sequentially.

Two transactions \emph{conflict} if they access the same storage
location, and at least one access is a write.
Transactions that do not conflict are said to \emph{commute},
because interleaving them in any order yields the same transaction and
storage states.
At a more granular level,
two bytecode operations \emph{conflict} if applying them in different orders yields
different storage states. Most bytecode operations, such as arithmetic operations
and others that interact with local state, commute with one another. Bytecode
operations that interact with shared state, on the other hand, can potentially
conflict.

The EVM operations \sload and \sstore read from and write to persistent storage, respectively,
and are used for nearly every state access and state modification.
By far the most common conflicts arise from conflicting
these two operations.
Other kinds of conflicts, while possible,
are assumed to be too rare to monitor.
For example,
one transaction might create a new contract,
while another calls the newly created contract in a way that creates a
race condition:
the call may or may not arrive before the contract is initialized. There is also a bytecode
operation that reads the a given account's balance, which is part of the blockchain's shared state.

We detect conflicts by associating a \emph{read-write lock} with each
storage location.
Each \var{SLOAD} (respectively, \var{SSTORE}) operation requests that location's
lock in read (write) mode.
If a transaction requests a lock that is already held in a conflicting
mode by another transaction,
the requesting transaction is rolled back and deferred to the next phase.
No locks are released until the concurrent phase ends, even those held
by aborted transactions. This is to ensure that no interleaving of
transactions in concurrent bin result in a conflict when re-executed
by validators.

\subsection{Sampling and Evaluation}
The Ethereum blockchain over the range of dates studied is much too
large to analyze in detail in a reasonable amount of time,
using reasonable computational resources.
Instead,
we chose to focus on seven historical periods between July 2016
(after the DAO fork) and December 2017.
Each period spans roughly one week,
with consecutive periods separated by 11 weeks,
ensuring that each day of the month is evenly sampled.
Due to computational limitations,
we analyzed every 10th block in each of those seven historical periods.

Our principal figure of merit is \emph{speed-up}:
the ratio between the time to execute a block's transactions sequentially,
and the time to execute the same transactions concurrently and
speculatively.
For now,
we use cumulative gas costs as a proxy for time,
measuring the ratio between
(1) the gas consumed to execute a block's transactions sequentially,
versus
(2) the maximum gas used by any thread in the parallel phase
plus the gas needed to execute the sequential phase.
Note that a conflicted transaction may be counted twice,
once in the concurrent phase, and once in the sequential phase.

\subsection{Baseline Results}

\begin{figure}[H]
\centering
\begin{subfigure}{.45\textwidth}
  \centering
  \includegraphics[scale=0.5]{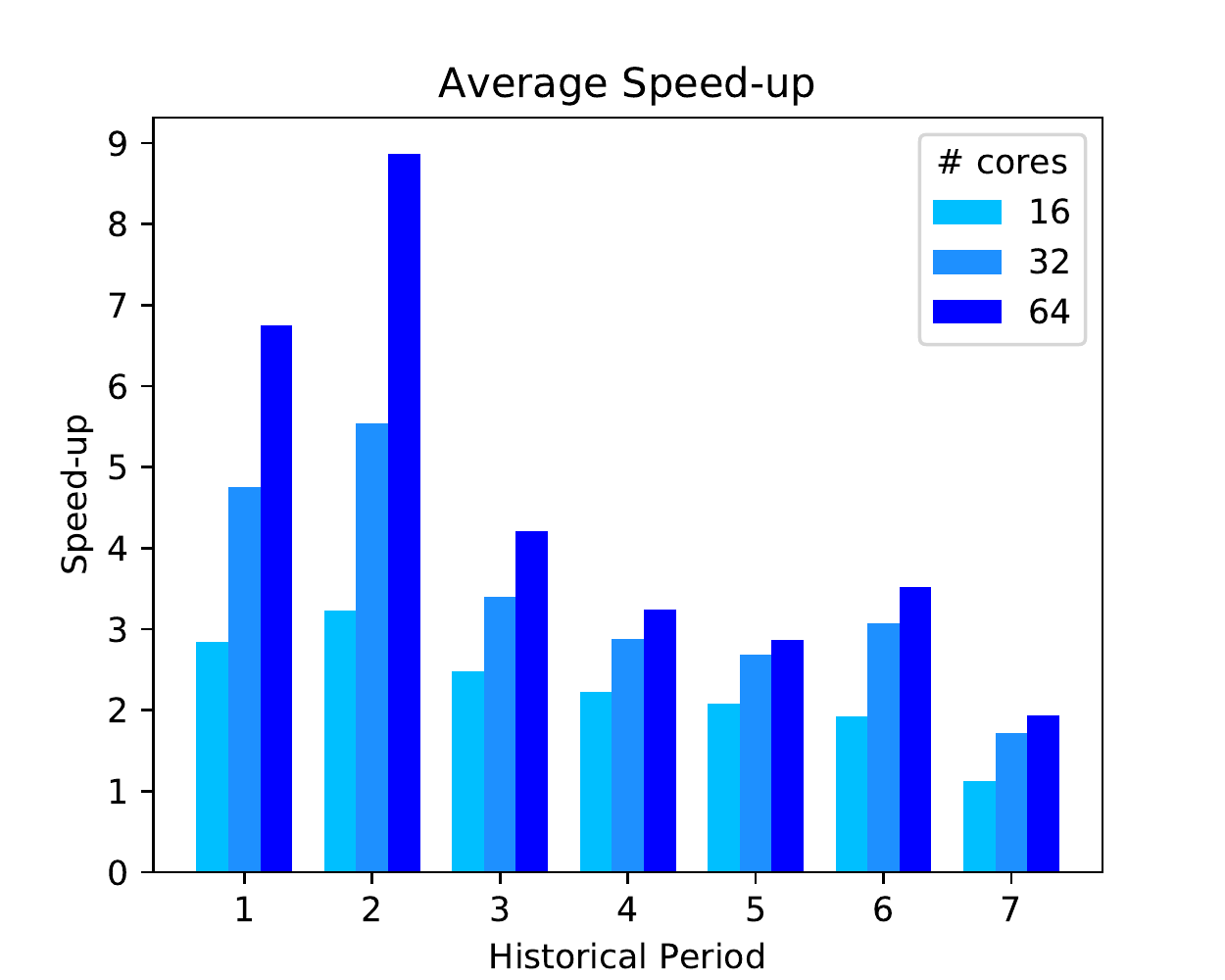}
  \caption{Average speed-up}
  \label{fig:avg_speedup}
\end{subfigure} \hfill
\begin{subfigure}{.45\textwidth}
  \centering
  \includegraphics[scale=0.5]{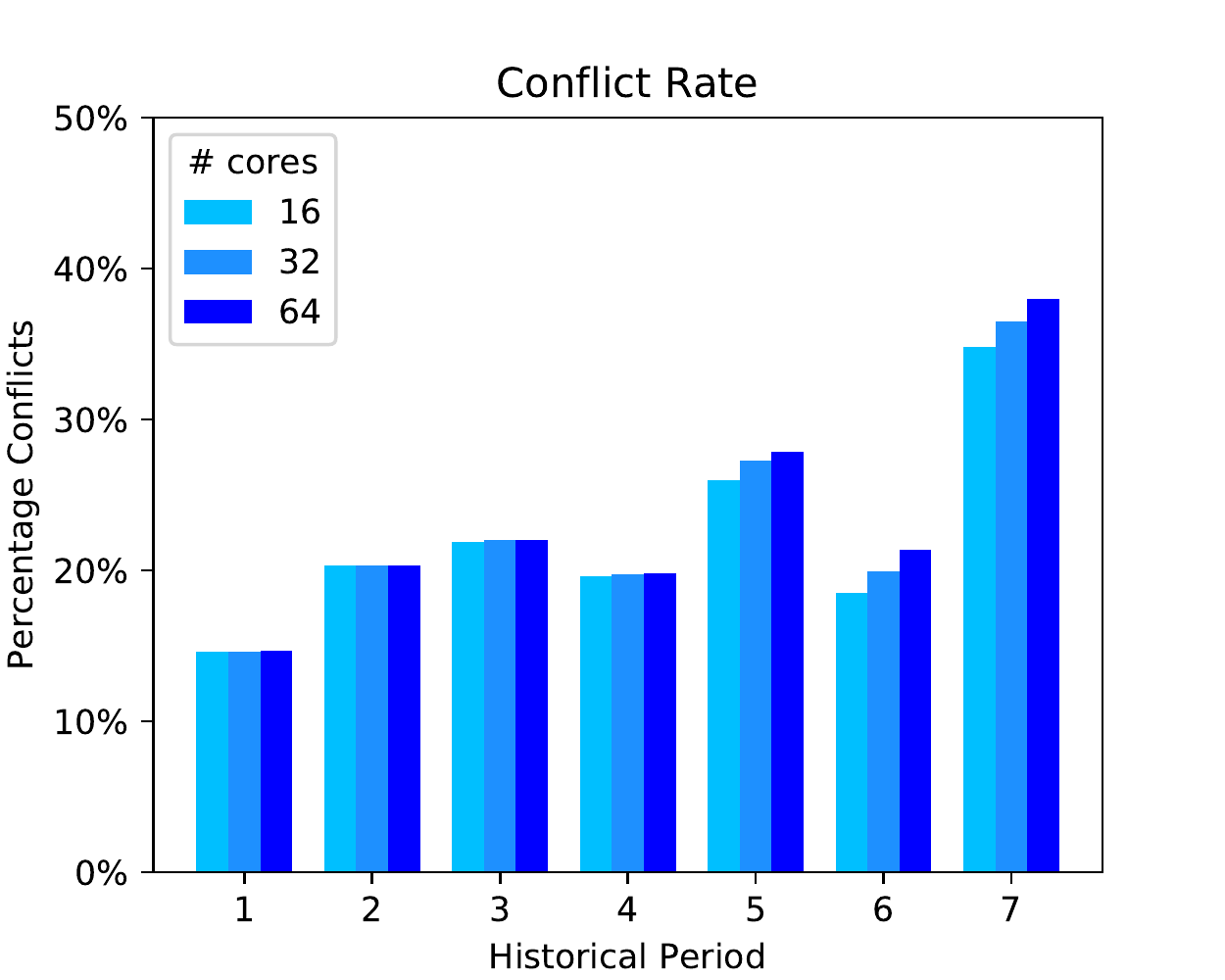}
  \caption{Average conflict rate}
  \label{fig:avg_aborts}
\end{subfigure} \hfill
\caption{Average speed-up and conflict rate for each of the seven historical time intervals. Increased Ethereum activity over time is reflected by decreasing speed-up and higher conflict rates.}
\label{summary}
\end{figure}

Figure \ref{summary} summarizes speed-up statistics over the seven
historical periods, along with the \emph{conflict rates},
or the percentage of contract calls that abort per block.
The average speed-up and conflict rate are shown for simulated VMs
of 16, 32, and 64 cores,
where averages are weighted by the number of contract call
transactions in each block. 

Earlier periods display higher speed-ups and lower conflict rates
because transaction volume and contention are low.
For example,
speed-up is as high as 3.23 in the second historical period on a
simulated EVM with 16 cores,
and this number rises to 8.87 for a 64 core EVM.
During the same interval,
contract calls abort at a rate of only 20\%. 

As the volume of transactions increases over time, however,
so does the rate of transaction conflict,
so more time is spent sequentially re-executing transactions aborted
during the concurrent phase. This naturally leads to lower speed-ups, since
there is less opportunity to parallelize transaction execution.
Indeed, during the December 2017 period,
with 16 threads, roughly 34\% of transactions abort.
Nevertheless, it is notable that there is still a modest but positive
speed-up of 1.13 even then.
Moreover, this speed-up effectively doubles,
to slightly more than 2 when there are 64 threads.
When transaction volume is higher, using more threads yields more speed-up.

\subsection{Speed-up Distributions}

The average speed-up of each historical period provides little insight into how the blocks' speed-ups are distributed in each period. Here, we further analyze the performance of speculative execution by looking at the distributions of these speed-ups. Due to space limitations, we focus on historical periods 5, 6, and 7, since these have the highest transaction volumes. 

\begin{figure}[H]
\centering
\begin{subfigure}{.3\textwidth}
  \includegraphics[scale=0.4]{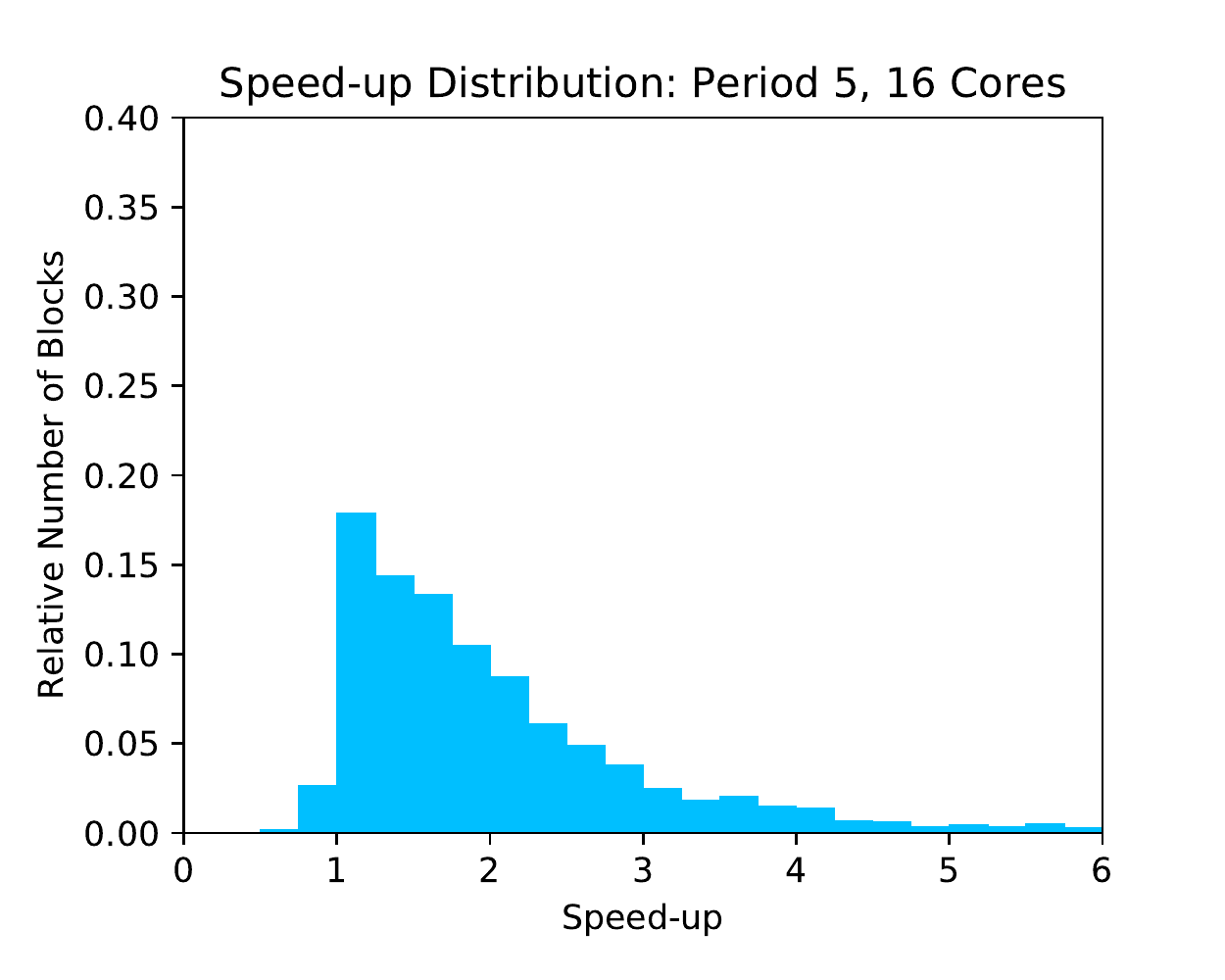}
\end{subfigure} \hfill
\begin{subfigure}{.3\textwidth}
  \includegraphics[scale=0.4]{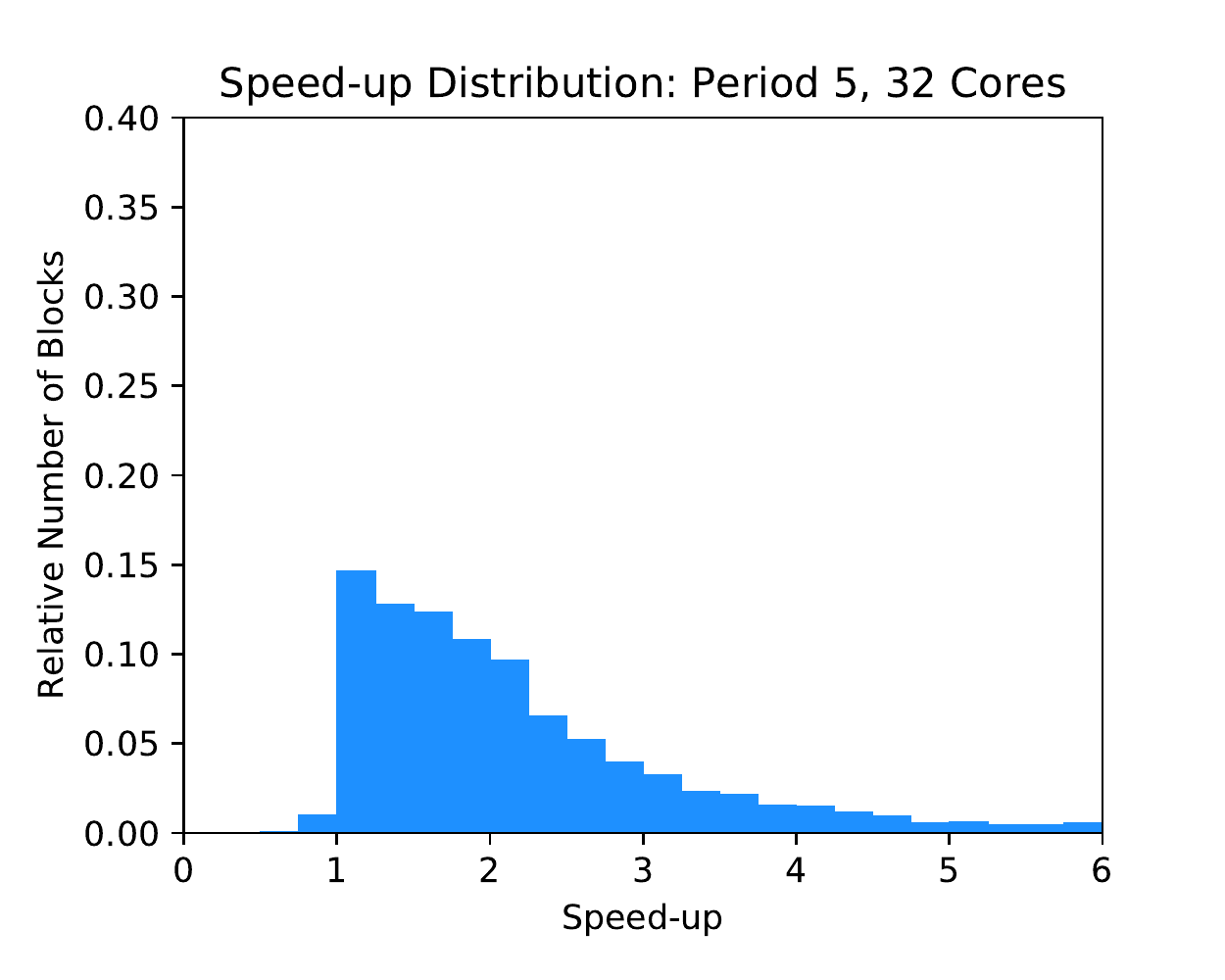}
\end{subfigure} \hfill
\begin{subfigure}{.3\textwidth}
  \includegraphics[scale=0.4]{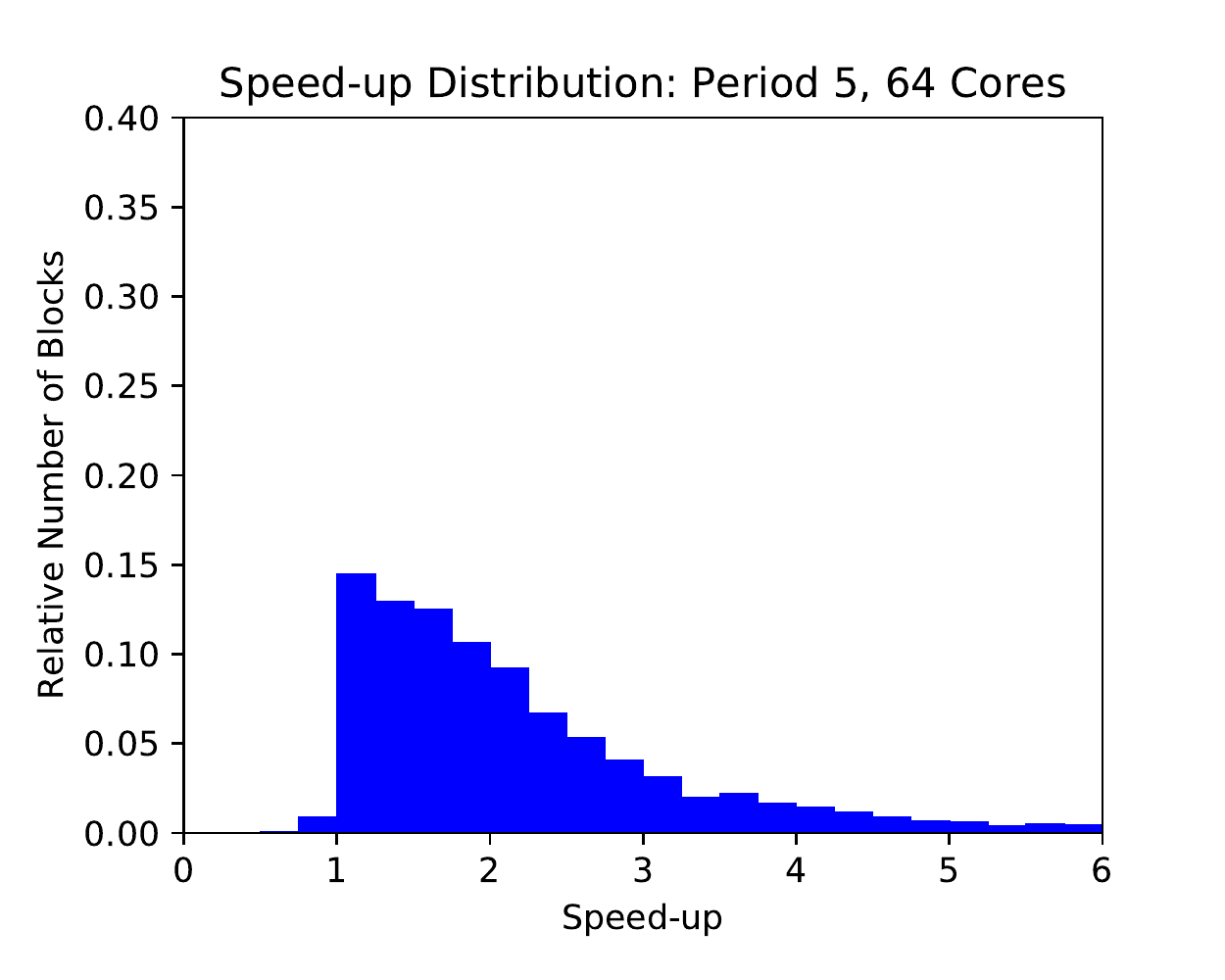}
\end{subfigure} \hfill

\begin{subfigure}{.3\textwidth}
  \includegraphics[scale=0.4]{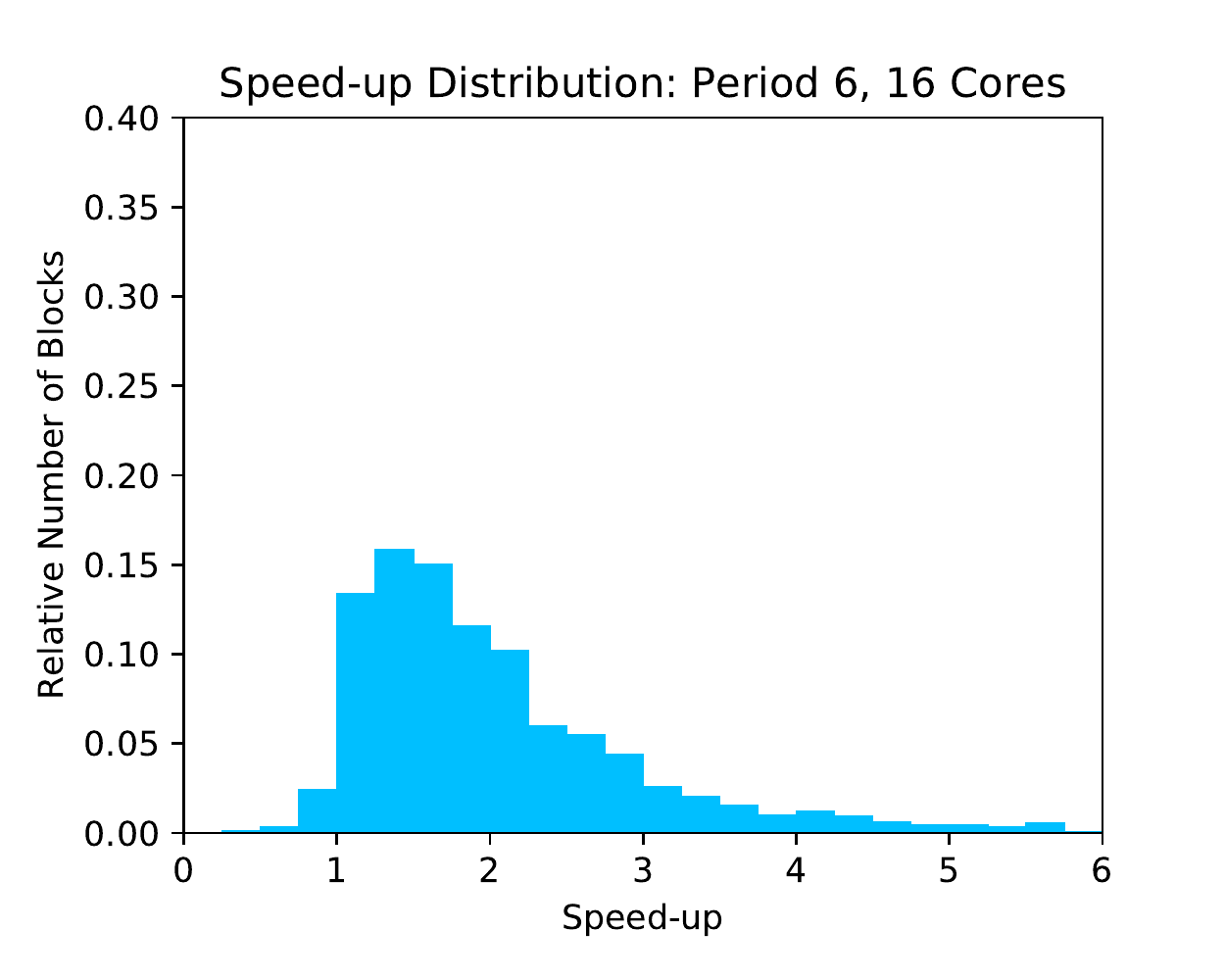}
\end{subfigure} \hfill
\begin{subfigure}{.3\textwidth}
  \includegraphics[scale=0.4]{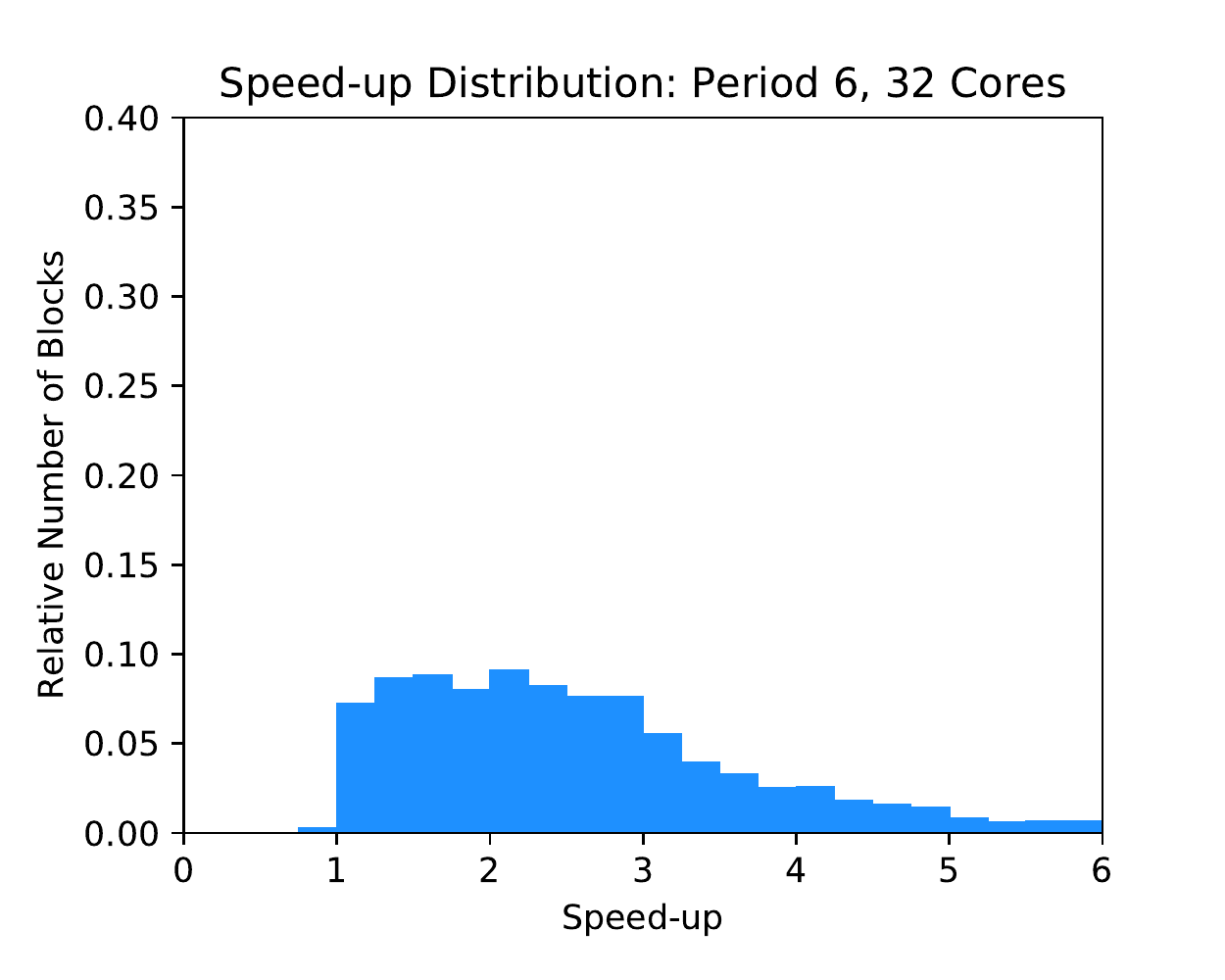}
\end{subfigure} \hfill
\begin{subfigure}{.3\textwidth}
  \includegraphics[scale=0.4]{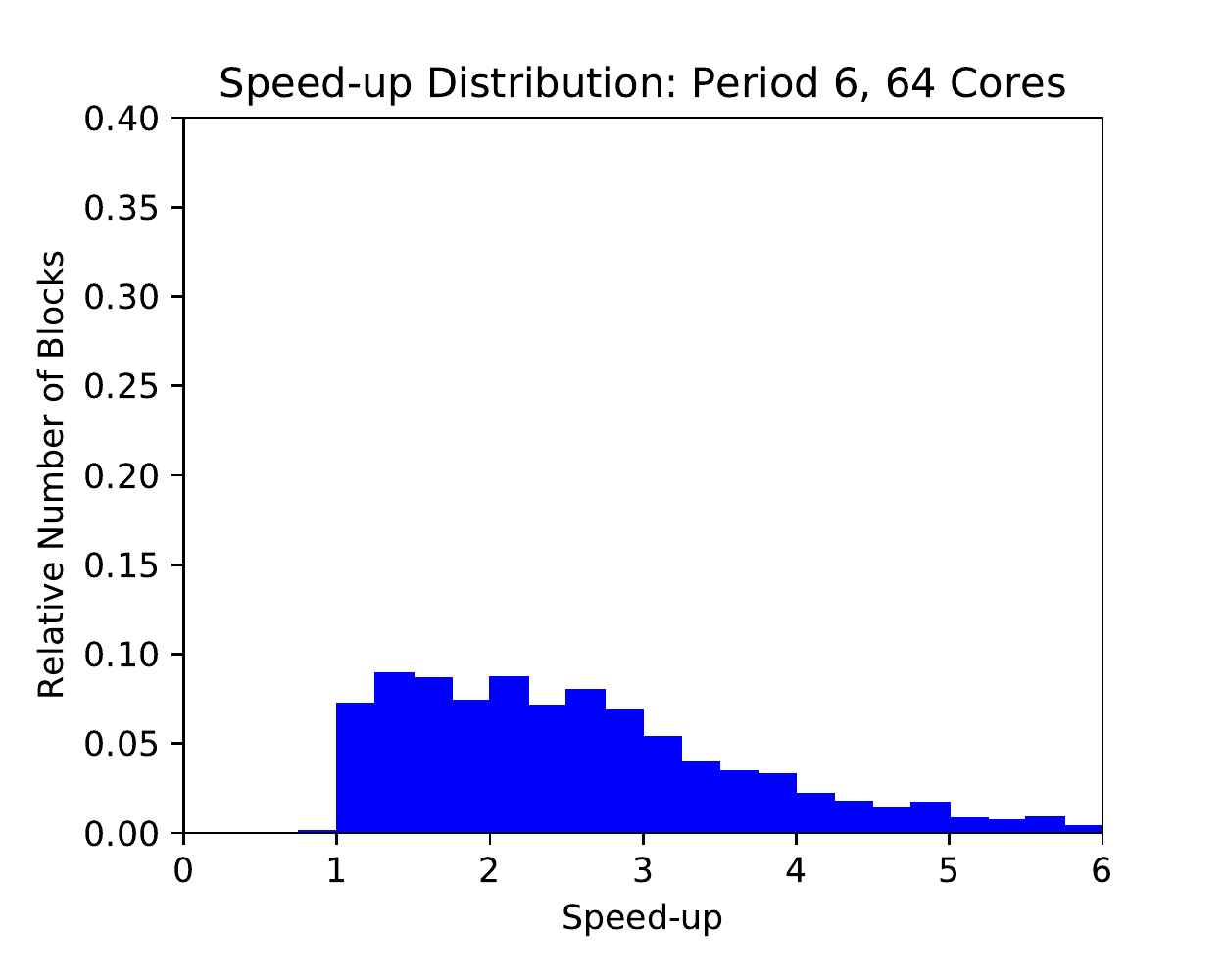}
\end{subfigure} \hfill

\begin{subfigure}{.3\textwidth}
  \includegraphics[scale=0.4]{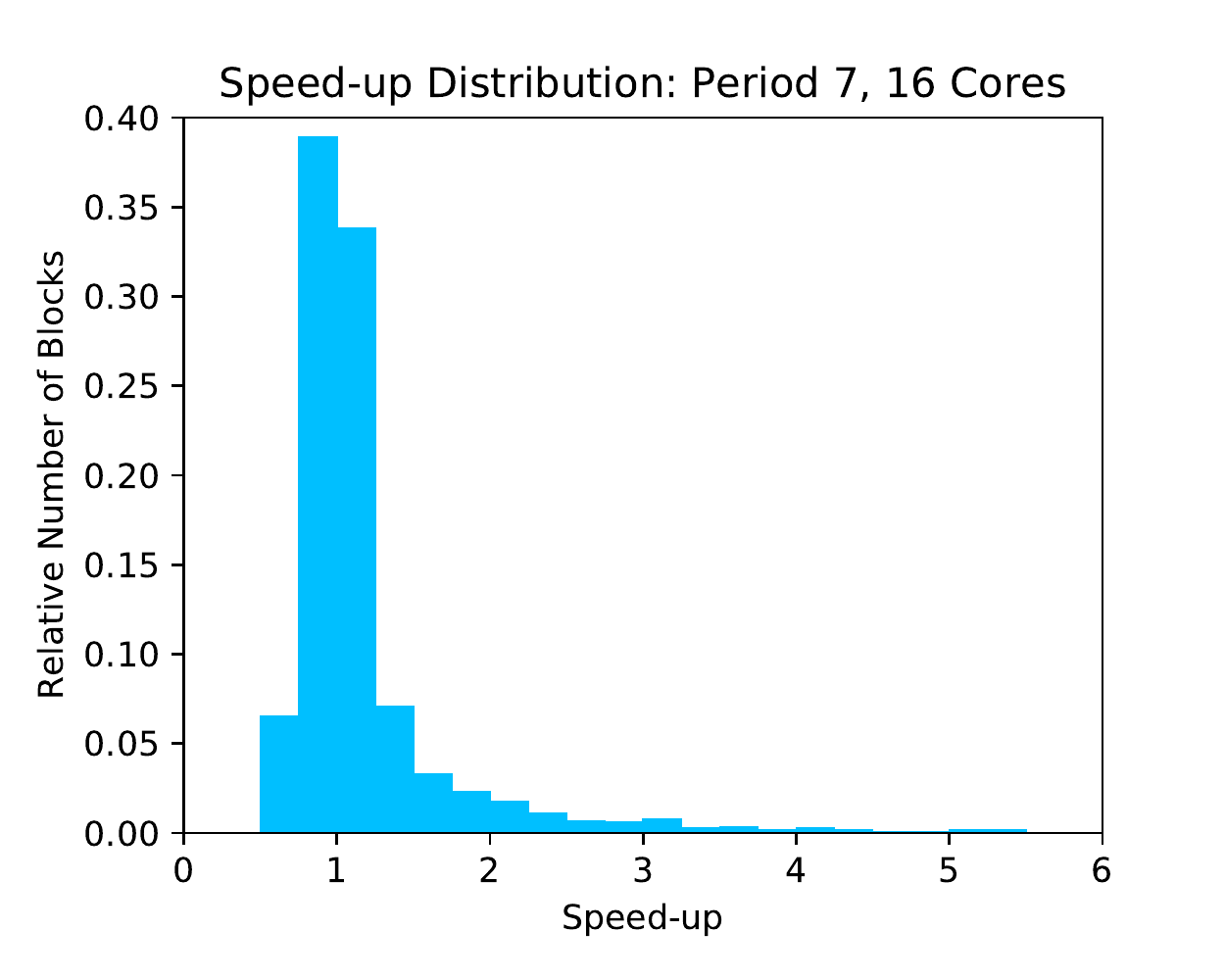}
\end{subfigure} \hfill
\begin{subfigure}{.3\textwidth}
  \includegraphics[scale=0.4]{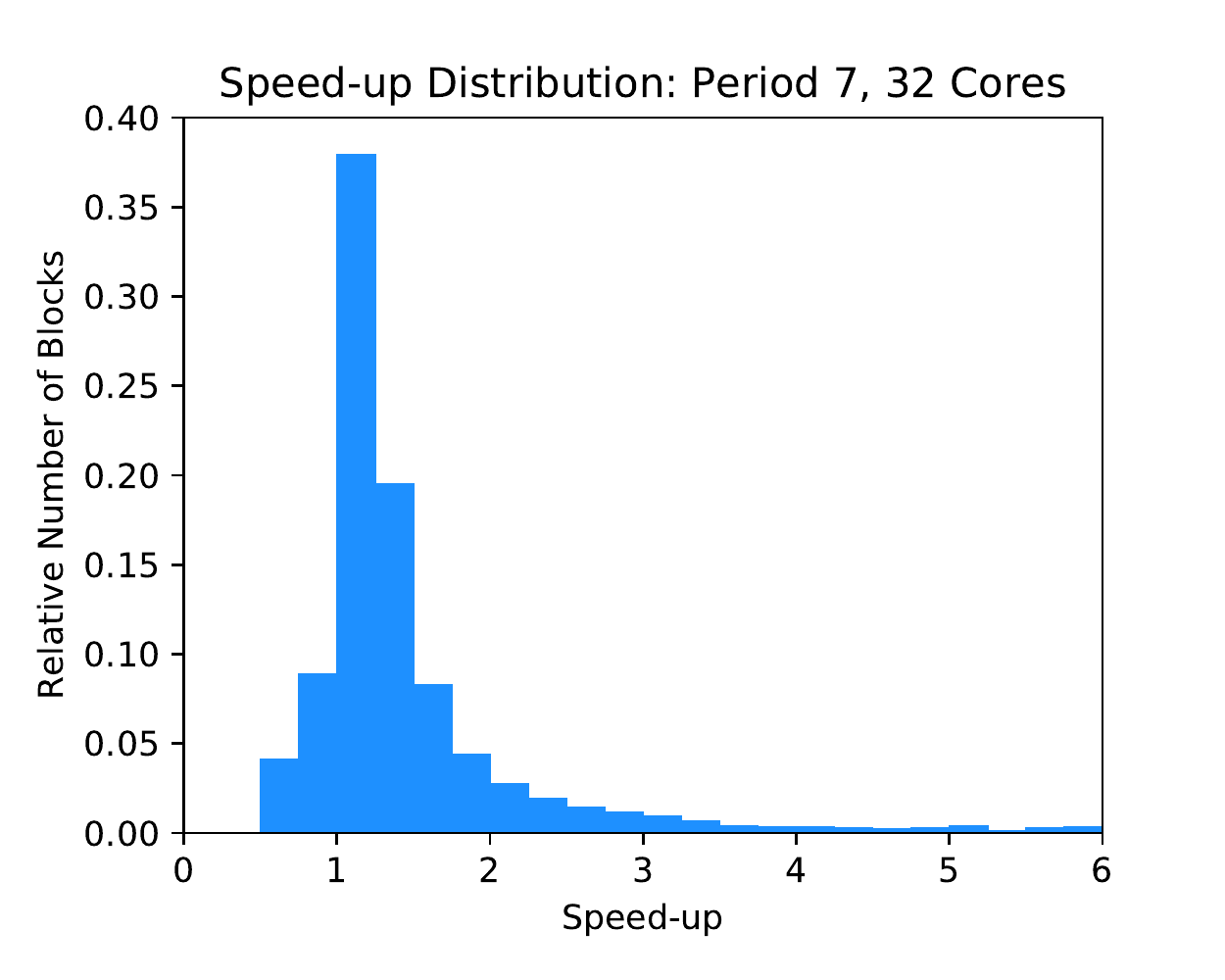}
\end{subfigure} \hfill
\begin{subfigure}{.3\textwidth}
  \includegraphics[scale=0.4]{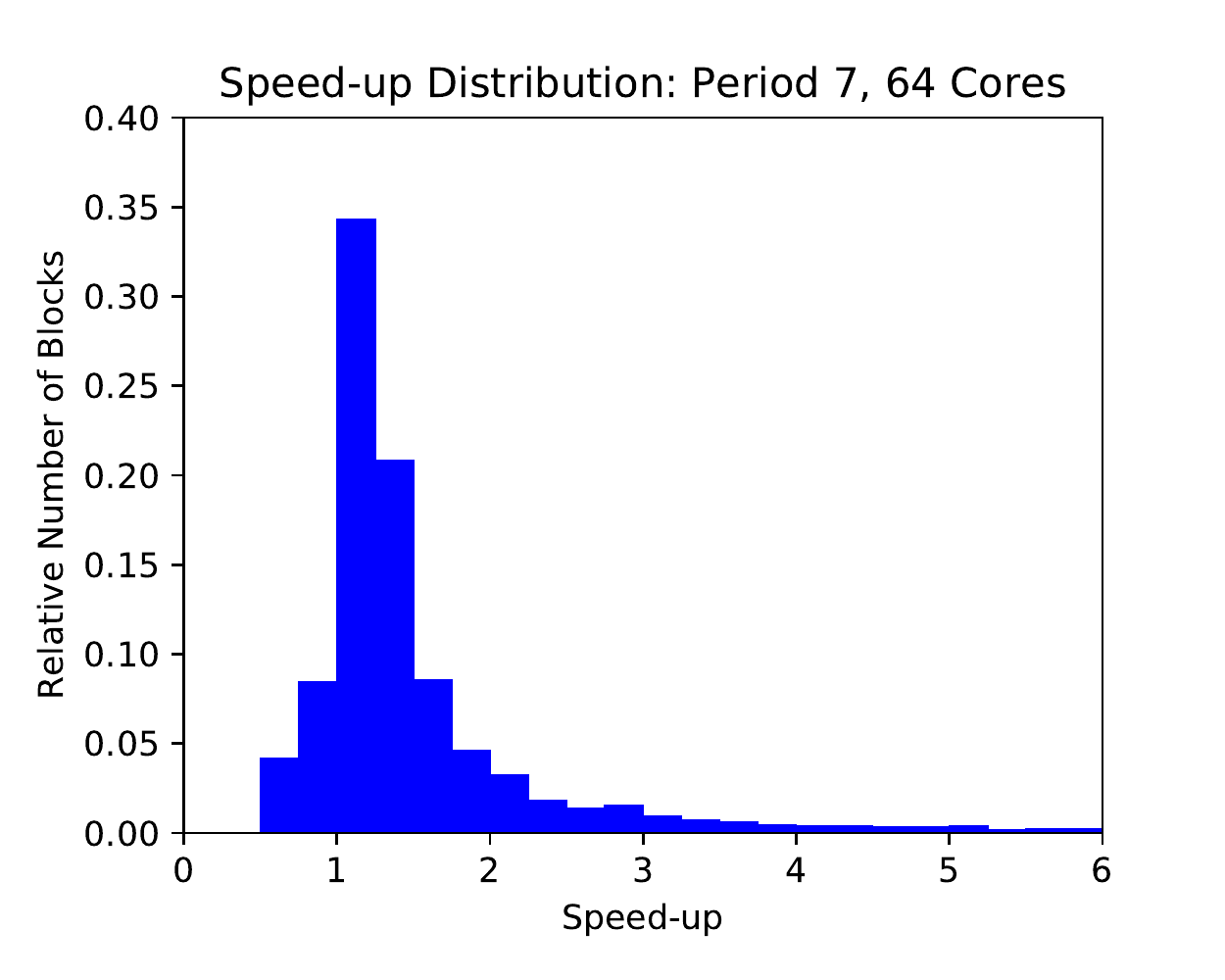}
\end{subfigure} \hfill
\caption{Speed-up distributions for each combination of historical periods 5, 6, and 7, and with 16, 32, and 64 cores. Each plot depicts one such combination. Blocks are binned in increments of 0.25 according to speed-up realized. Distributions are normalized so that total area is roughly 1. }
\label{distros}
\end{figure}

Speed-up distributions of the three periods are depicted in Figure~\ref{distros}. Each plot is normalized to represent an
approximate probability density, so that the y-axis is interpreted as a relative count.
Since there is relatively less contention in periods 5 and 6, their speed-up distributions are flatter, with
longer tails. This is because more blocks during these periods are able to exhibit a higher amount of speed-up.
However, when the transaction rate is higher,
as in seventh period of December 2017,
the distributions are concentrated closed to 1,
since more work is wasted as the number of aborted transactions rises.  

As noted before, some blocks may realize a slowdown when using our
concurrent execution strategy,
a result of the cost of aborting and re-executing conflicting transactions.
The relative number of such blocks is shown in plots of Figure~\ref{distros},
where regions to the left of $1$ represent blocks that are slowed down.
Making more cores available to the EVM appears to reduce the number of slowed-down blocks
for each of the three historical periods.
Though fewer than 3\% blocks are slowed down
for both periods 5 and 6 when using 16 cores, this percentage jumps to
46\% for Period 7.
However, when the EVM's number of cores is increased to 32,
the percentage of blocks that are slowed down drops substantially to only 13\%.

\subsection{Storage Hot-Spots}

Next, we investigate transaction contention by analyzing how often certain storage addresses are accessed by transactions.
In particular,
we look at memory addresses that are conflict points for pairs of transactions,
and how many times each such address results in a conflict.
Addresses that attract a high number of conflicts are informally called \emph{hot-spots}.

\begin{figure}[H]
\centering
\begin{subfigure}{.3\textwidth}
  \centering
  \includegraphics[scale=0.4]{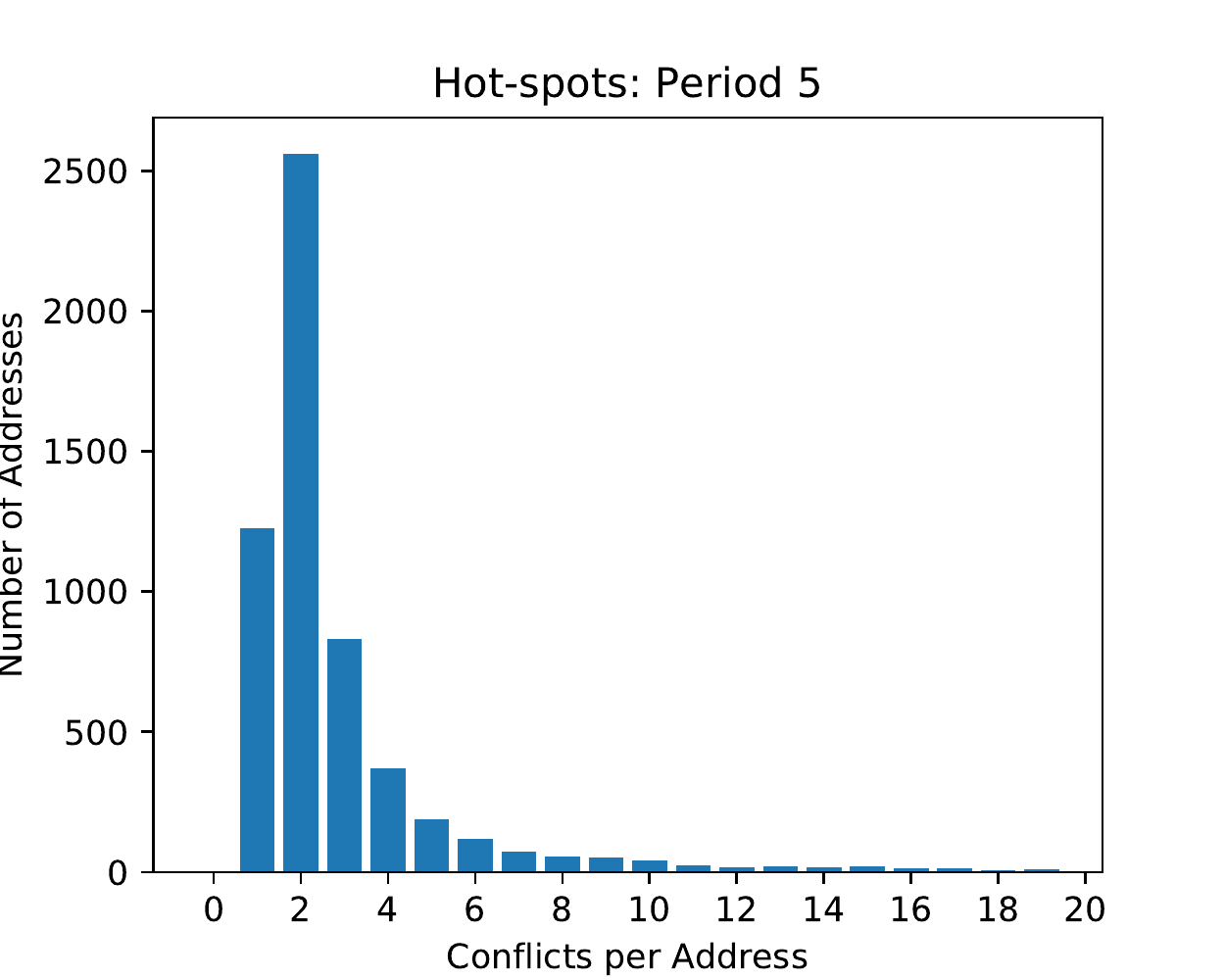}
  \label{fig:hotspot1}
\end{subfigure} \hfill
\begin{subfigure}{.3\textwidth}
  \centering
  \includegraphics[scale=0.4]{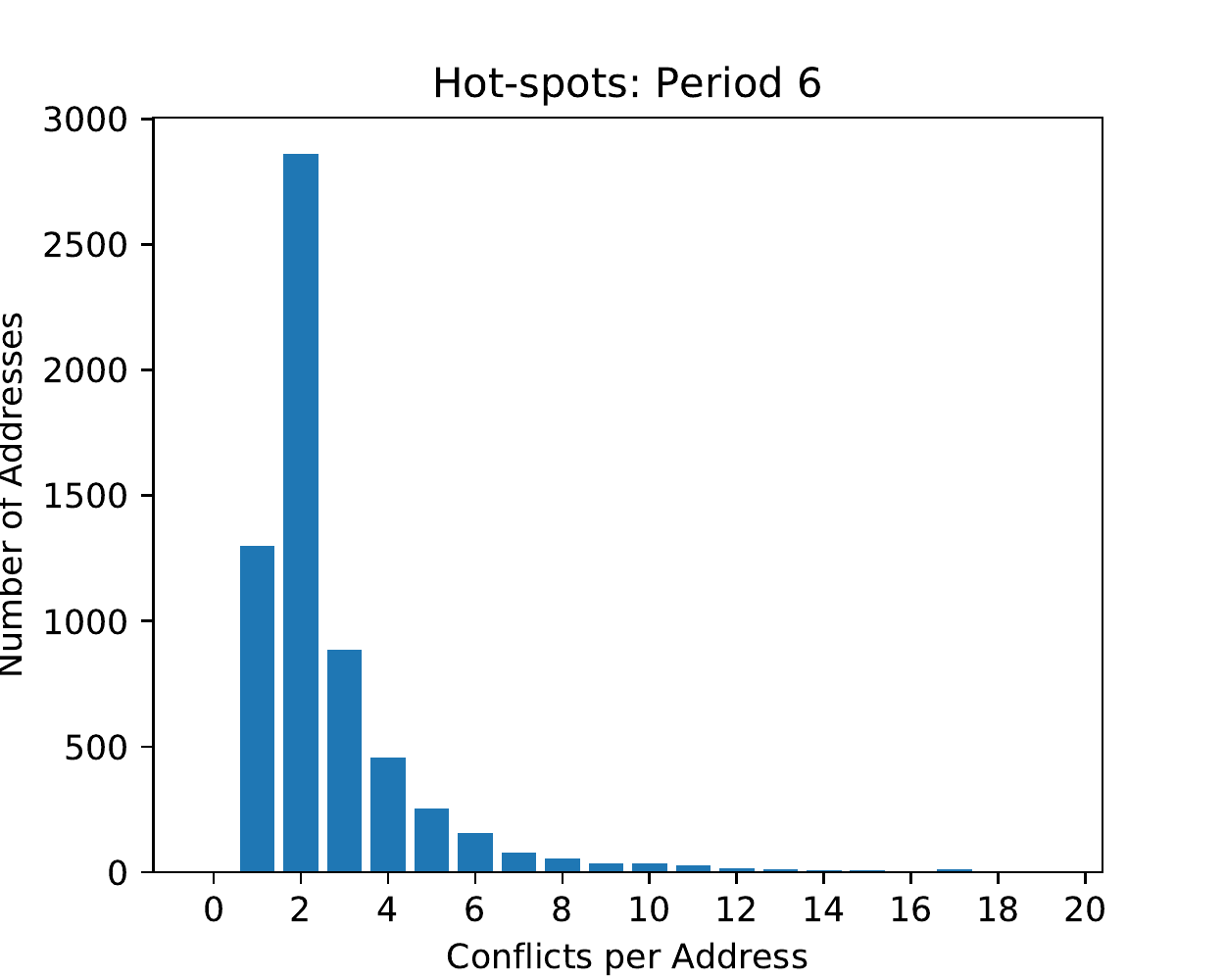}
  \label{fig:hotspot2}
\end{subfigure} \hfill
\begin{subfigure}{.3\textwidth}
  \centering
  \includegraphics[scale=0.4]{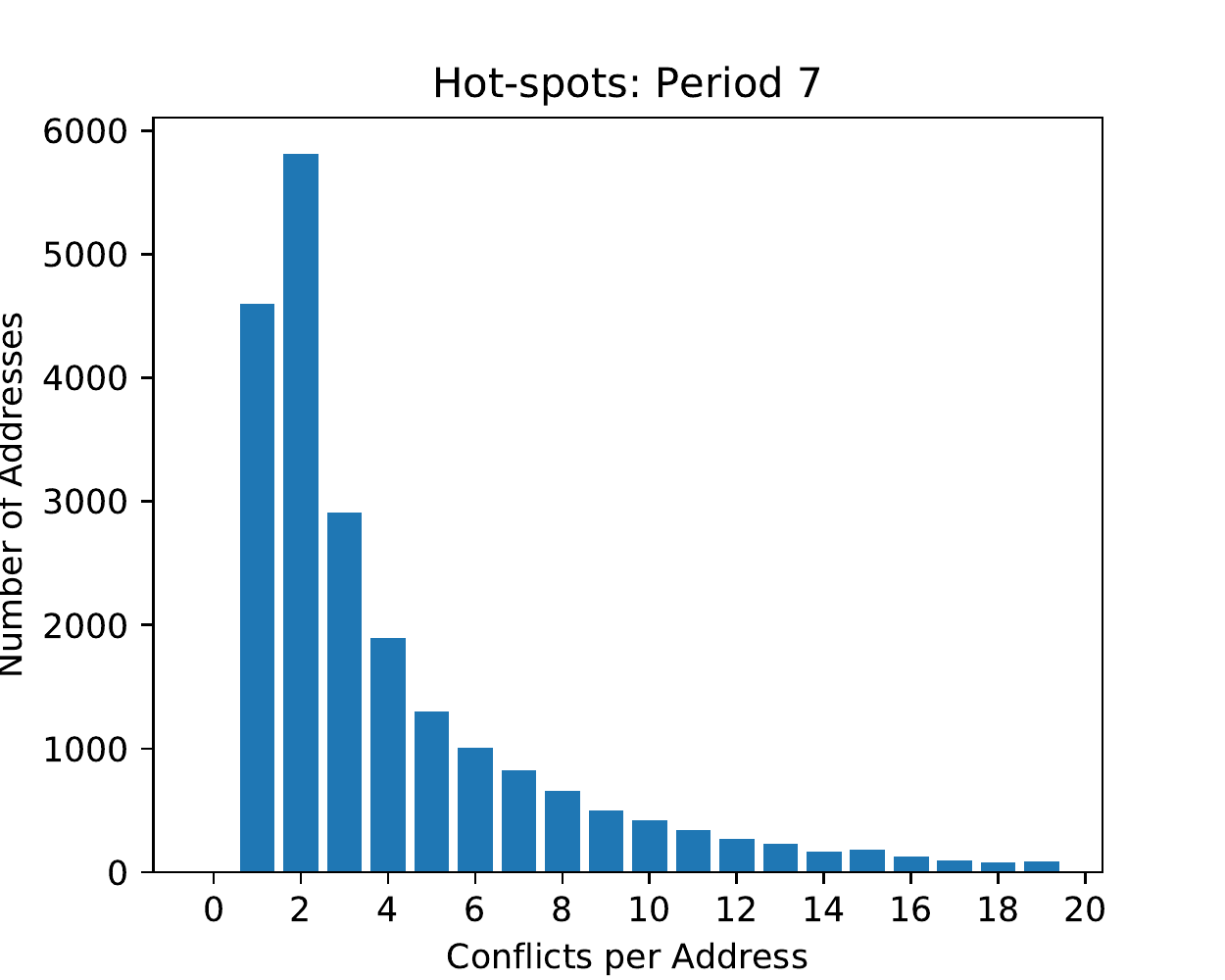}
  \label{fig:hotspot3}
\end{subfigure} \hfill

\caption{Histograms of conflicts per address. Addresses that results in a large number of conflicts are hot-spots. Period 7 has a much heavier tail, corresponding to many more address hot-spots.}
\label{hotspots}
\end{figure}

In Figure \ref{hotspots}, we plot histograms illustrating the number of conflicts per address, for historical periods 5, 6, and 7. Each storage address with at least one conflict is binned according to how many conflicts occurred at that address. Periods 5 and 6 have relatively few storage addresses with high contention, and this pattern is similar to the earlier historical periods. However, Period 7's
histogram has a much heavier tail than the other two histograms, meaning that there are many more storage addresses with larger
numbers of conflicts. In other words, there is a handful of addresses that many different transactions attempt to access, most likely
a result of so many transactions calling the same few contracts.
We elaborate on this observations in Figure \ref{hotspot} below.


%

%

%% file: alternative.tex
\section{Alternative Experiments}
\label{alternative}
We extended the baseline experiment with a number of other experiments
intended to test the effectiveness of alternative strategies,
and to test the sensitivity of our approximations.

\subsection{Sampling}

\begin{figure}[H]
  \centering
  \renewcommand{\arraystretch}{1.2}
  \begin{tabular}{|l|c|c|c|c|c|c|}
    \hline
    \multirow{2}{3cm}{\textbf{Sampling rate}} & \multicolumn{3}{c|}{\textbf{Speed-up}} & \multicolumn{3}{c|}{\textbf{Abort rate}} \\
    \cline{2-7}
    & \textbf{16 cores} & \textbf{32 cores} & \textbf{64 cores} & \textbf{16 cores} & \textbf{32 cores} & \textbf{64 cores} \\
    \hline
    1-in-10 sampling & 2.063 & 2.694 & 2.859 & 25.98\% & 27.23\% & 27.88\% \\ \hline
    All blocks & 2.085 & 2.717 & 2.913 & 25.96\% & 27.21\% & 27.86\% \\ \hline
  \end{tabular}
  \caption{Accuracy of 1-in-10 sampling for historical period 5.}
\end{figure}


\subsection{Multiple Phases}
\begin{wrapfigure}{r}{0.4\textwidth}
\centering
\includegraphics[scale=0.5]{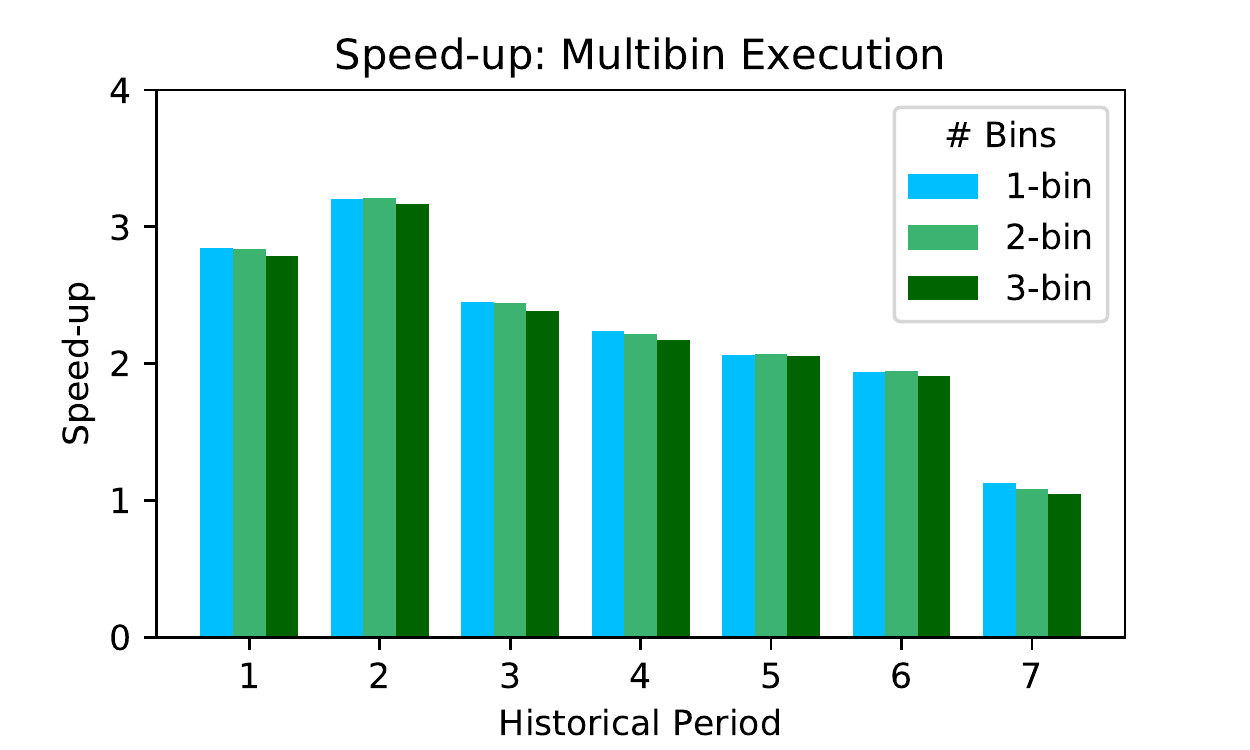}
\caption{Speed-up with multiple concurrent phases for transaction execution.}
\label{multibin}
\end{wrapfigure}
The greedy, two-phase strategy can be generalized to encompass
multiple concurrent phases, so that
each transaction that was deferred in one concurrent phase
is instead re-executed in another subsequent concurrent phase.
It is possible that multiple concurrent phases could provide additional speed-up.
However, as illustrated in Figure \ref{multibin}, we found in our experiments that executing two concurrent phases
almost always yields less speed-up than executing a single concurrent
phase.
This decrease is due to the duplicate work performed by transactions
rolled back in multiple phases, with not enough additional speed-up yielded
in the latter concurrent phases. Therefore, in practice, one current phase is sufficient to
realize almost all potential concurrency.

\subsection{Data Conflicts}
\begin{wrapfigure}{r}{0.4\textwidth}
\centering
\includegraphics[scale=0.5]{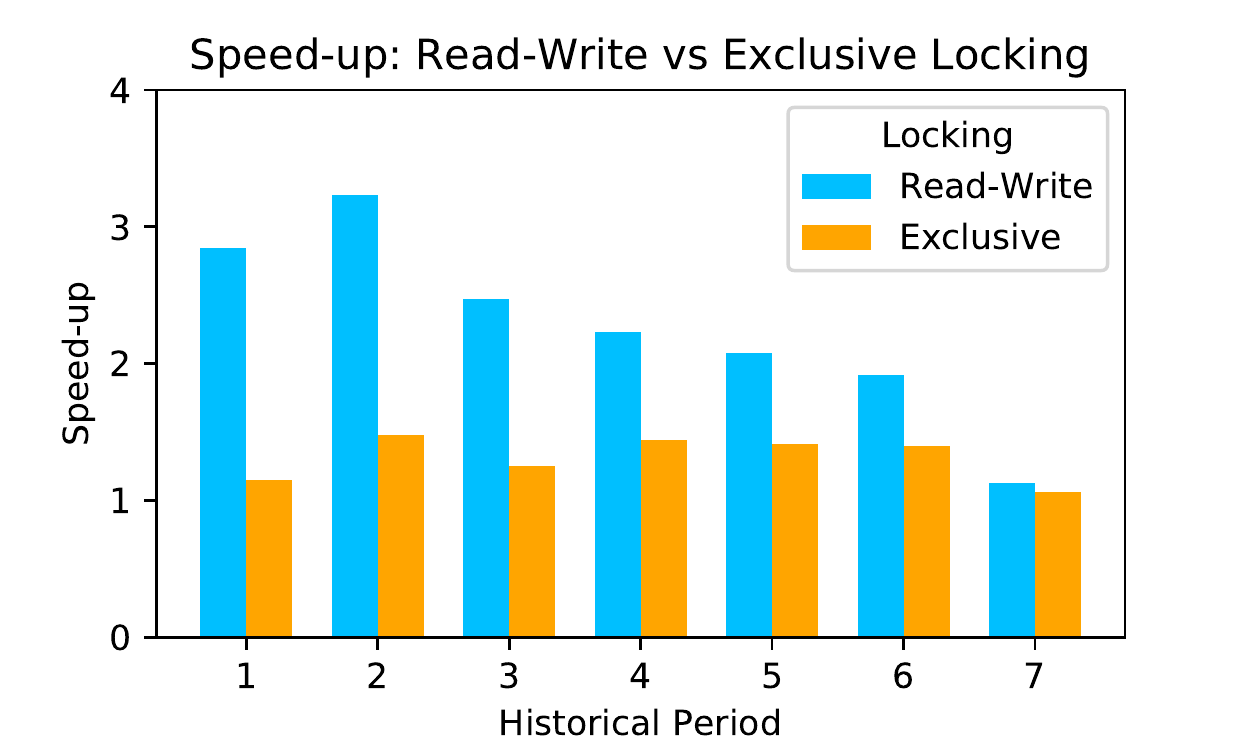}
\caption{Speed-up when the EVM uses mutex locks to lock storage addresses.}
\label{locks}
\end{wrapfigure}
In our simulated concurrent EVM, transactions access storage addresses by first
acquiring a read-write lock. This allows multiple transactions to read an address,
with no conflict, if there are no concurrent writers. In principle, this decreases the
number of conflicts when compared to using mutex locks, hence increasing the potential speed-up.

To determine whether read-write locks reduce data conflicts in practice, 
we investigated the effect of simplifying the conflict model
by merging all data accesses into a single conflict set, by using mutex locks for conflict detection.
With the exception of the last historical period in which speed-up is already low, the speed-ups were substantially
worse than the speed-ups obtained by distinguishing between read and write accesses. See Figure \ref{locks} for
a comparison of the two locking schemes when using 16 cores.
These results suggest that there is significant value in implementing
a concurrent EVM with read-write locks, instead of the simpler mutex locks.

\subsection{Proxies for Time}
\begin{wrapfigure}{r}{0.4\textwidth}
\centering
\includegraphics[scale=0.5]{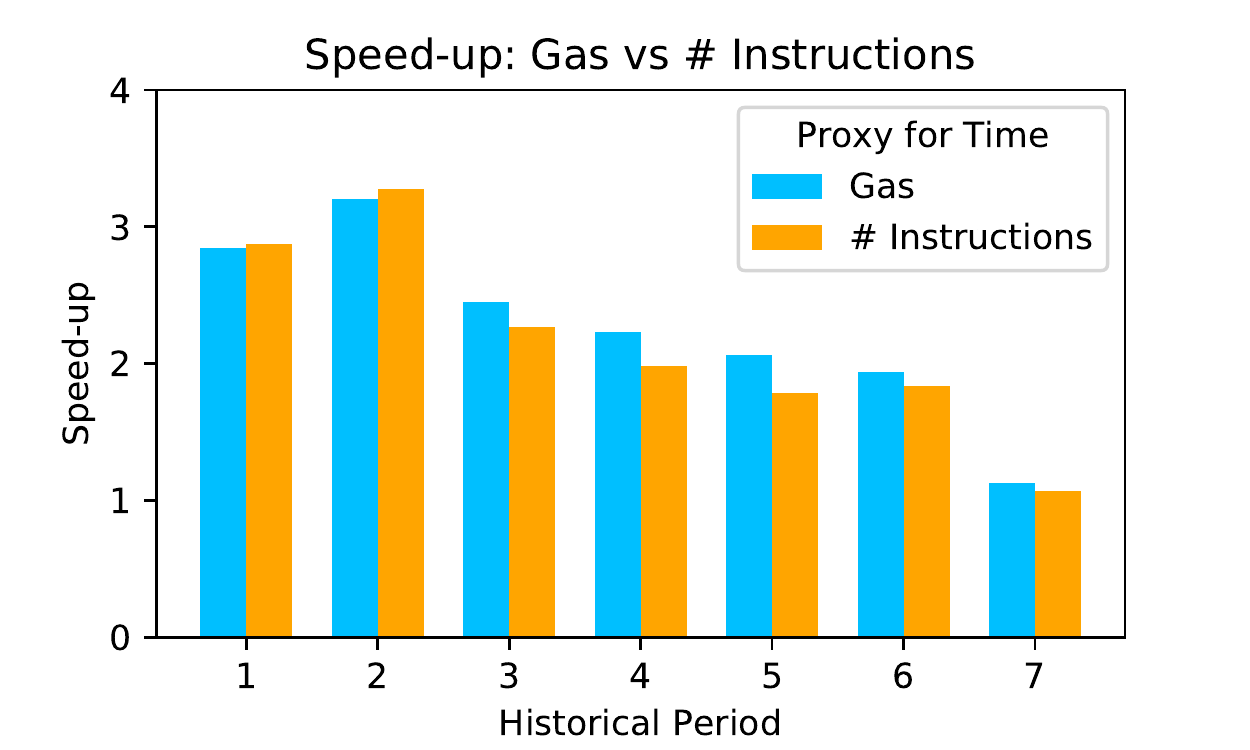}
\caption{Speed-up when instruction count is used to measure transaction execution time.}
\label{proxies}
\end{wrapfigure}

Since we do not have access to an actual concurrent EVM,
we must estimate how long it takes to execute each transaction.
There are two straightforward choices:
we can count the number of instructions executed by each transaction,
or we can tally the gas cost of executing the transaction's instructions.
The first choice assumes that each EVM instruction takes roughly the same
time to execute,
while the second assumes that instruction gas cost is roughly proportional to
execution time.
For example, the
arithmetic operations \emul and \ediv require 5 units of gas,
while an \sstore instruction can cost as much as 20000 units of gas.

All speed-ups reported so far were measured in terms of gas costs.
As a sanity check, for 16 cores,
we recomputed speed-ups using instruction count as a proxy for time.
These speed-ups are shown in Figure \ref{proxies}. There does not appear
to be any significant qualitative difference between gas cost and instruction count.

\subsection{Static Conflict Prediction}
\begin{wrapfigure}{r}{0.4\textwidth}
\centering
\includegraphics[scale=0.5]{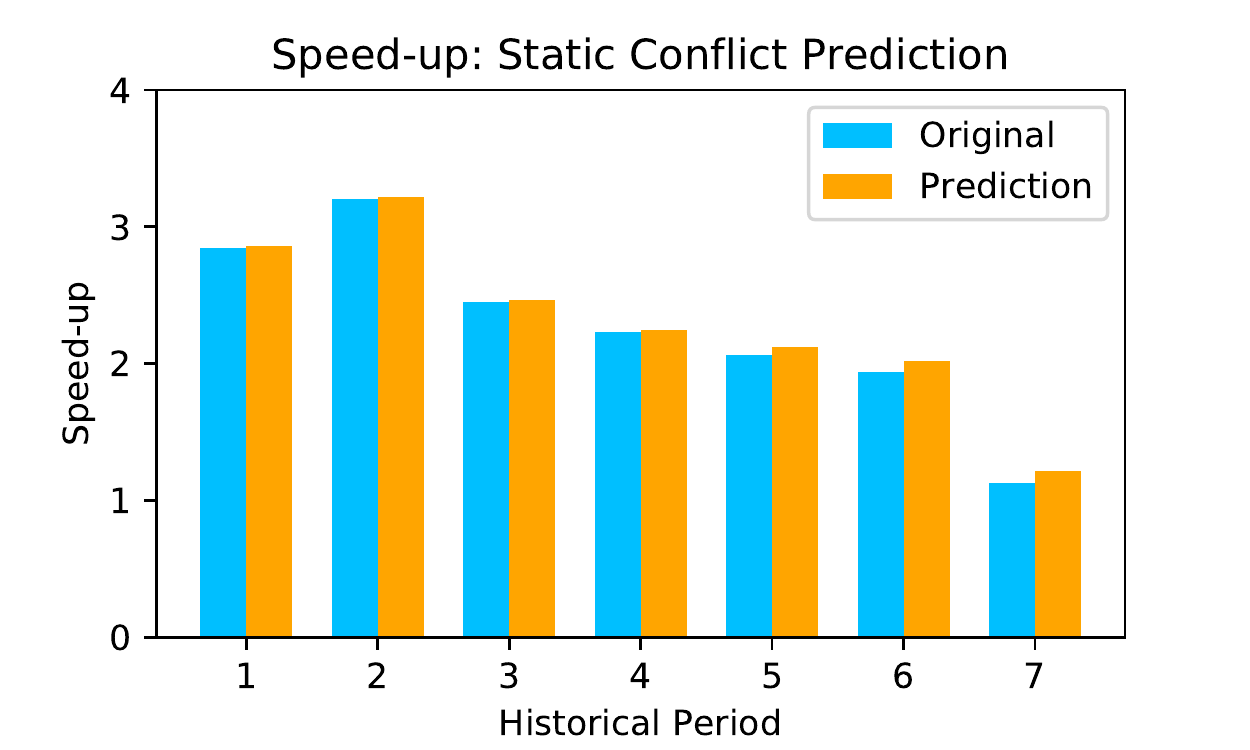}
\caption{Speed-up under static conflict prediction. In effect, aborted transactions costs
can be ignored.}
\end{wrapfigure}

If we were able to predict whether a transaction would abort if
executed speculatively,
then we could save the cost of transaction roll-back and retry.
We can simulate the effect of a perfect conflict predictor simply by
ignoring the cost of aborted transactions.
However, doing so yield a negligible change in speed-ups,
less than $0.1\%$ in most cases.
The only exception is the very last period (December 2017),
where contention was very high.
For that historical period,
the average speed-up increases from $1.13$ to $1.22$.
These numbers suggest that static conflict analysis,
if it accurate enough,
may be yield modest gains during periods of high contention.

\subsection{Omitting Hot-spot Contracts}
\label{hotspot}

In most of the previous analysis, period 7 stands out among the selected historical periods as being particularly
high volume and contentious. Recall that this period was sampled from December 2017, which is close to peak
Ethereum transaction activity. More specifically, period 7 occurred when there was great interest from the general
public over \emph{CryptoKitties} \cite{bowles_2017} \cite{Cryptokitties}, a recreational game deployed on Ethereum in which users create, breed, and trade virtual cats. The popularity of CryptoKitties is especially apparent when using any blockchain explorer to browse Ethereum transactions during December 2017.
The popularity of CryptoKitties was responsible for congesting the Ethereum network, though interest in it has since died down.

In light of the CryptoKitties frenzy, we reanalyzed period 7 under a hypothetical scenario in which the CryptoKitties contracts
\footnote{The vast majority of calls to CryptoKitties are to its core contract (which has address hash
\texttt{0x06012c8cf97bead5deae237070f9587f8e7a266d}), and to an auction contract (which has address hash \texttt{0xb1690c08e213a35ed9bab7b318de14420fb57d8c}).}
did not exist. This is easily accomplished by ignoring all calls to the CryptoKitties contracts when tracing each block's transactions
and replaying them, and calculating the resulting statistics. 

\begin{figure}[H]
\centering
\begin{subfigure}{.45\textwidth}
  \centering
  \includegraphics[scale=0.5]{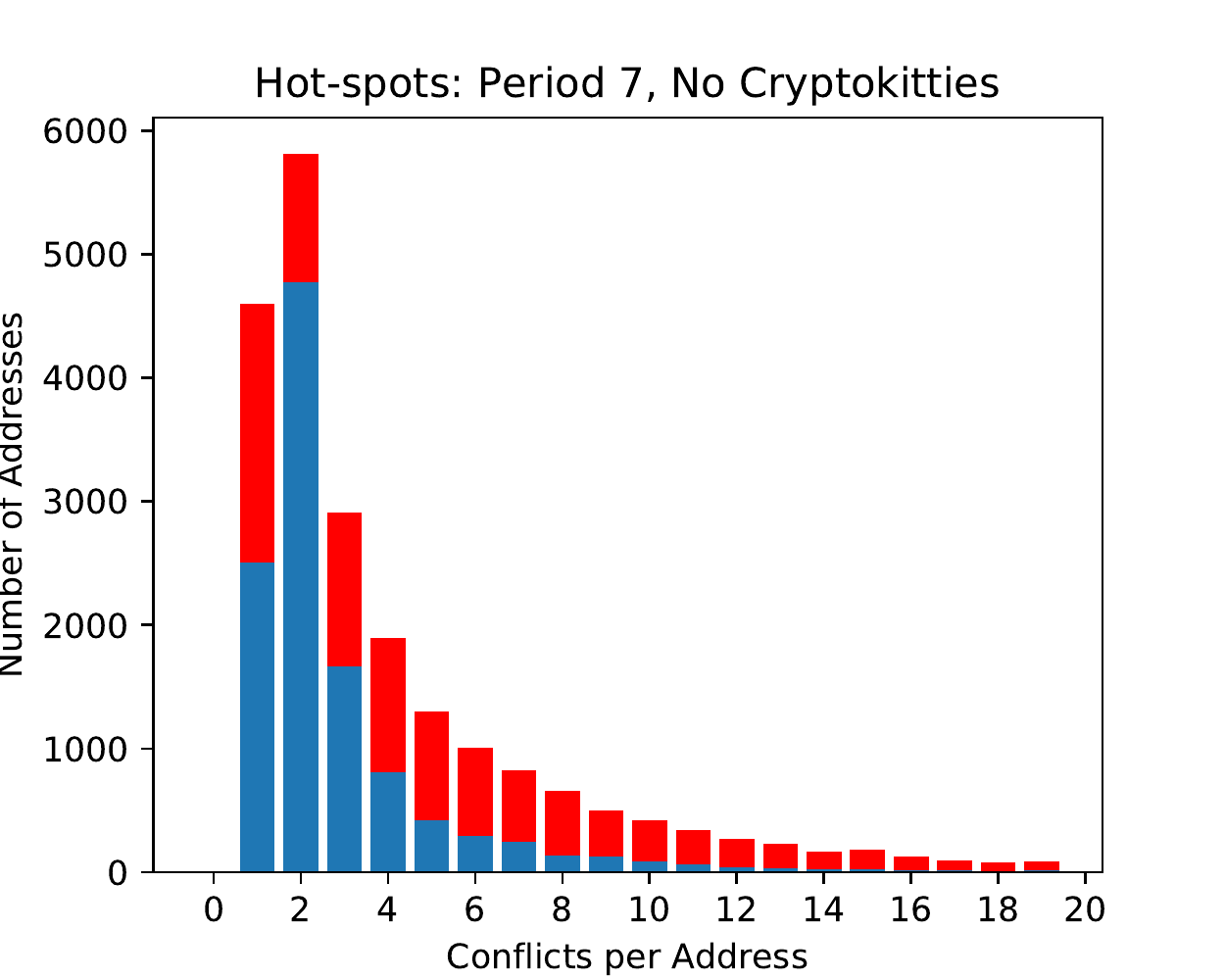}
  \caption{The effect on hot-spot counts when CryptoKitties contract calls are ignored.}
  \label{nocrypto}
\end{subfigure} \hfill
\begin{subfigure}{.45\textwidth}
  \centering
  \includegraphics[scale=0.5]{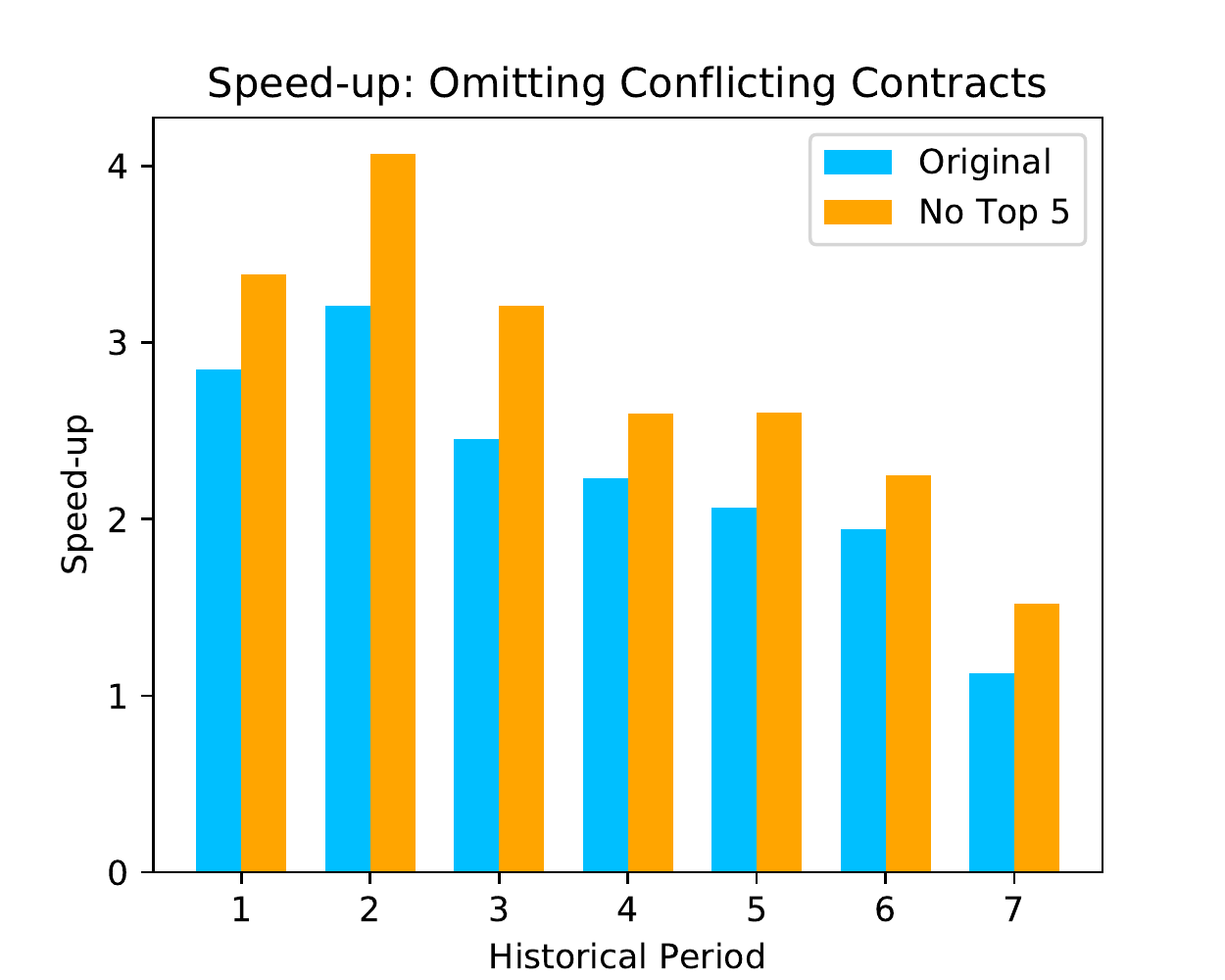}
  \caption{Speed-up when calls to the five most conflicting contracts are ignored.}
  \label{notop5}
\end{subfigure} \hfill
\caption{Speed-up analysis after removing high-contention or high-volume contracts.}
\label{filter}
\end{figure}

After filtering out all calls to CryptoKitties contracts, speed-up using 16 cores rises from 1.13 to 1.65. These calls accounted for about 31\% of all contract calls, and furthermore, by filtering out CryptoKitties, the number of contract calls per block drops from 62 to 43, which is much closer to the 46 calls per block in period 6. While a speed-up of 1.65 does not quite match speed-up from older periods, it is still clear from this simple analysis that much of the contention in period 7 is caused solely by CryptoKitties.

To further illustrate the impact that CryptoKitties had, we reproduce the conflict histogram from the previous section, using the same hypothetical scenario with no CryptoKitties. Recall that the conflict histogram of period 7 had a much heavier tail than the corresponding
histograms of older intervals. However when CryptoKitties is removed, period 7's histogram (Figure \ref{nocrypto}) has a much thinner tail, resembling the other histograms. Indeed, in period 6, 10\% of conflicting storage addresses have at least 5 conflicts, but in period 7, this same number is 30\%. However, if CryptoKitties contracts are ignored, then only 14\% or conflicting address have at least 5 conflicts. These numbers demonstrate that CryptoKitties is responsible for many storage hot-spots.

We generalize this analysis to the other historical periods by determining the top five most conflicting smart contracts in each period, and reproducing scenarios where none of the highly conflicting contracts were called. As shown in Figure \ref{notop5}, this results in a noticeable speed-up in each period, not just the 7th. Most of these highly conflicting contracts are actually token contracts; in fact, the most contentious contract from each period is either a token or a token exchange contract. Therefore, analyzing these small sets of contracts may provide insight into how to reduce contention when speculatively executing smart contracts in parallel.

%% file: discussion.tex
\section{Discussion}
As noted, this study is exploratory,
incorporating various approximations and omissions.
Most such omissions are the result of the absence of a standard
concurrent EVM implementation.
This study does not account for some EVM overheads,
such as the costs for \emph{value transfers},
where one account transfers ether directly to another,
without modifying storage.
Value transfers typically commute, and are likely much faster to execute
than contract calls, though they do make up the majority of all Ethereum transactions.
Other sources of overhead include solving a cryptographic puzzle to compute proof of work,
which would affect miners' speed-up but not the validators'.

In the absence of a timing model for a concurrent EVM,
this study uses gas costs (or instruction counts)
as proxies for time when computing speed-up.
As noted, both proxies yield essentially equivalent results. 

This study relies on sampled blocks because the volume of data in the
Ethereum blockchain is simply too large to make exhaustive analysis
practical or rewarding. An archive synchronization of the blockchain
was a major computational overhead for this study, in addition to
recovering transaction traces. These overheads may be reduced 
as further Ethereum tools and utilities are developed.

There are several reasons why speculative parallelism may 
sometimes yield little, or even negative speed-up.
For example,
a block might contain one transaction substantially longer than the others,
whose execution time dominates the block execution time. In this case,
it is impossible to achieve much speed-up, no matter how these transactions
are scheduled and distributed among multiple cores.
Or if a block contains very few contract call transactions, there is little opportunity for speed-up.
If a block's transactions all access the same storage location,
perhaps because they all access the same popular
contract~\cite{Cryptokitties},
then speculation will produce a negative speed-up 
as a result of the cost of rolling back so many misspeculated transactions.



%% file: conclusion.tex
\section{Conclusions}
Our results suggest that a simple speculative strategy based on
read-write set overlap can produce non-trivial speed-ups,
but that such speed-ups will decline as transaction rates and conflict rates increase.
More aggressive strategies,
such as adding additional parallel phases,
seem to provide little additional benefit,
because conflict appears to be bursty:
if one transaction conflicts with another,
then it probably conflicts with multiple others.

The results of this study suggest that 
the most promising way to further increase parallelism in
Ethereum-style smart contract execution is to reduce the conflict rate,
perhaps by focusing on reducing unnecessary conflicts.
We observed that splitting transactions' data sets into
read sets and write sets decreased conflict rates substantially,
suggesting that conflict rates are sensitive to the
\emph{semantics} of concurrent operations on shared data.
This observation suggests that conflict rates might be reduced even
further if the execution engine could do a better job of recognizing
when operations commute at the semantic level.
For example,
transactions that increment or decrement the same account balance
(a common occurrence) have overlapping read and write sets,
and are therefore deemed to conflict.
At the semantic level, however,
these operations commute (in the absence of overflow or underflow),
so as long as the virtual machine's memory operations are atomic,
those operations need not conflict.
(Our study could not detect which conflicts are real, and which are artifacts,
because only compiled bytecode was available for analysis).

It might be profitable to investigate the effects of endowing the
virtual machine with intrinsic data types such as atomic counters or
atomic sets that provide many commuting mutator operations.
Studying highly conflicting token contracts may provide insight
into which kinds of data types or operations would best alleviate
contention.

Periods of high contention and low speed-up are caused by a relatively
small number of popular contracts.
Currently, smart contract designers have no guidance how to avoid
speculative data conflicts,
nor any incentive to do so.
Our results suggest that there is a need to devise incentives for
smart contract programmers to design contracts in ways that reduce conflicts,
either by eliminating spurious conflicts,
or by exploiting improved commuting bytecode instructions.